\documentclass[12pt]{article}
\usepackage[a4paper,left=2cm,right=1.2cm,top=3cm,bottom=4cm]
{geometry}
\usepackage{amsmath,amstext, amsthm, empheq, extpfeil,  
amsbsy,amssymb,marvosym,fancyhdr,graphicx,amscd,amsfonts,latexsym,delarray,stackrel,lineno,color,cite,appendix,xcolor,url,hyperref}
\usepackage{wasysym}
\input xypic
\usepackage{srcltx}
\usepackage{mathrsfs}
\usepackage{float} 
\usepackage{caption}

\usepackage[all]{xy}
 \definecolor{trust}{rgb}{0,1,1}

\definecolor{grey}{rgb}{0.95,0.95,0.95}

\newtheorem{thm}{Theorem}[section]

\newtheorem{lem}[thm]{Lemma}

\newtheorem{defn}[thm]{Definition}

\def\qqq{\,,\quad \forall}

\def\pert{{\rm Pert}}

\def\Out{{\rm Out}}

\def\Aut{{\rm Aut}}

\def\Tr{{\rm Tr}}
\def\tr{{\rm tr}}

\def\C{{\mathbb C}}

\def\N{{\mathbb N}}

\def\R{{\mathbb R}}

\def\Z{{\mathbb Z}}
\def\H{{\mathbb H}}

\def\Tr{{\rm Tr}}
\def\tr{{\rm tr}}

\def\cA{{\mathcal A}}
\def\cB{{\mathcal B}}

\def\cH{{\mathcal H}}

\def\part{\partial}

\def\part{\partial}

\newcommand{\ie}{{\it i.e.\/}\ }

\newcommand{\cf}{{\it cf.\/}\ }

\def\Int{{\mbox{Int}}}

\def\H{{\mathbb H}}

\definecolor{trust}{rgb}{0,1,1}

\def\qqq{\,,\quad \forall}

\begin{document}
\title{Geometry and the Quantum}
\author{Alain Connes}
\maketitle
\tableofcontents 
\section{Introduction}
The ideas of noncommutative geometry are deeply rooted in both physics, with the predominant influence of the discovery of Quantum Mechanics, and in mathematics where it emerged from the great variety of examples of ``noncommutative spaces" \ie of geometric spaces which are best encoded algebraically  by a  noncommutative algebra.

It is an honor  to present an overview of the state of the art of the interplay of noncommutative geometry with physics on the occasion of the celebration of the centenary of Hilbert's work on the foundations of  physics.
Indeed, the ideas which I will  explain, those of noncommutative geometry (NCG) in relation to our model of space-time, owe a lot to Hilbert and this is so in two respects. First of course by the fundamental role of Hilbert space in the formalism of Quantum Mechanics as formalized by von Neumann, see \S \ref{spec sect}. But also because,  
as explained in details in \cite{corry,KS}, one can consider Hilbert to be   the first person to have speculated about a unified theory of electromagnetism and gravitation, we come to this point soon in \S \ref{sect hilbert}.

\subsection{The spectral point of view}\label{spec sect}
At the beginning of the eighties, motivated by the exploration of the many new spaces whose algebraic incarnation is noncommutative, I introduced a new paradigm, of spectral nature, for geometric spaces. It is based on the Hilbert space formalism of Quantum Mechanics and on mathematical ideas coming from $K$-theory and index theory. A geometry is given by a ``spectral triple" $(\cA,\cH,D)$ which consists of an involutive algebra $\cA$ concretely represented as an algebra of operators in a Hilbert space $\cH$ and of a (generally unbounded) self-adjoint operator $D$ acting on the same Hilbert space $\cH$. The main conceptual motivation came from the work of Atiyah and Singer on the index theorem and their realization that the Hilbert space formalism was the proper setting for ``abstract elliptic operators" \cite{Atiyah}.

 To fix ideas: a compact spin Riemannian manifold is encoded as a spectral triple by letting the algebra of functions act in the Hilbert space of spinors while the Dirac operator $D$ plays the role of the inverse line element, as we shall amply explain below. But the key examples that showed, very early on, that the relevance of this new paradigm went far beyond the framework of Riemannian geometry comprised duals of discrete groups, leaf spaces of foliations and deformations of ordinary spaces such as the noncommutative tori which were themselves a prime example of noncommutative geometric spaces as shown in \cite{Co-CR}.
 
  In the middle of the eighties it became clear that the new paradigm of geometry, because of its flexibility, provided a new perspective on the geometric interpretation of the detailed  structure of the Standard model and of the Brout-Englert-Higgs mechanism. Over the years this new point of view has been considerably refined and is now able to account for the 
 extremely complicated Lagrangian of Einstein gravity coupled to the standard model of particle physics. It is obtained from the spectral action developed in our joint work with A. Chamseddine in \cite{cc2}. The spectral action is the only natural additive spectral invariant of a noncommutative geometry.
 
  The noncommutative geometry dictated by physics is the product of the ordinary $4$-dimensional continuum by a finite noncommutative geometry which appears naturally from the classification of finite geometries of $KO$-dimension equal to $6$ modulo $8$ (\cf \cite{cc5,mc2}). The compatibility of the model with the measured value of the Higgs mass was demonstrated in \cite{acresil} due to the role in the renormalization of the scalar field already present in  \cite{ac2010}.  In \cite{acmu1,acmu2}, with Chamseddine and Mukhanov, we gave the conceptual explanation of the finite noncommutative geometry from Clifford algebras and obtained a higher form of the Heisenberg commutation relations between $p$ and $q$, whose irreducible Hilbert space representations correspond to $4$-dimensional spin geometries. The role of $p$ is played by the 
Dirac operator and the role of $q$ by the Feynman slash of coordinates using Clifford algebras. The proof that all spin geometries are obtained relies on deep results of immersion theory and ramified coverings of the sphere.  The volume of the $4$-dimensional geometry is automatically quantized by the index theorem; and the  spectral model, taking into account the inner automorphisms due to the noncommutative nature of the Clifford algebras, gives Einstein gravity coupled with a slight extension of the standard model, which is a Pati-Salam model. This model was shown in our joint work with A. Chamseddine and W. van Suijlekom \cite{acpati1,acpati2} to yield unification of coupling constants.

\subsection{Gravity coupled with matter}\label{sect hilbert}
   
As explained in detail in \cite{corry}, one can consider Hilbert as the first to have fancied a unified theory of electromagnetism and gravitation.
According to \cite{corry}, in the course of pursuing this agenda, Hilbert  reversed his original idea of founding all of physics on electrodynamics, instead  treating the gravitational field equations as more fundamental. We have, in our investigations with Ali Chamseddine of the fine structure of space-time which is revealed by the Brout-Englert-Higgs mechanism, followed a parallel path: the starting point was that the NCG framework for geometry, by allowing to treat the discrete and the continuum on the same footing gives a clear geometric meaning to the Brout-Englert-Higgs sector of the Standard Model, as the signal of a discrete (but finite) component of the geometry of space-time appearing as a fine structure which refines the usual $4$-dimensional continuum.

 The action principle however was at the beginning of the theory still of traditional form (see \cite{Co-book}). In our joint work with Ali Chamseddine \cite{cc2} we understood that instead of imitating the traditional form of the Yang-Mills action, one could obtain the full package of the Einstein-Hilbert action\footnote{There is a well-known ``priority episode" between Hilbert and Einstein which is discussed in great detail in  \cite{corry,KS} and whose outcome, called the Einstein-Hilbert action, plays a key role in our approach} of gravity coupled with matter by a fundamental spectral principle. In the language of NCG this principle asserts that the action only depends upon the ``line element" \ie the inverse\footnote{In the orthogonal complement of its kernel} of the operator $D$. It follows then from elementary considerations of additivity for disjoint unions of spaces  that it must be of the form $\Tr(f(D/\Lambda))$ where $f$ is a function and $\Lambda$ is a parameter having the same dimension (that of an energy) as the inverse line element $D$. 

\subsection{Possible relevance for Quantum Gravity}\label{sect QG}

It will by now be clear to the reader that the point of view adopted in this essay is to try to understand from a mathematical perspective, how the perplexing combination of the Einstein-Hilbert action coupled with matter, with all the subtleties such as the Brout-Englert-Higgs sector, the V-A and the see-saw mechanisms etc..  can emerge from a simple geometric model. The new tool is the spectral paradigm and the new outcome is that geometry does emerge  on the stage where Quantum Mechanics happens, \ie  Hilbert space and linear operators. 

The idea that group representations as operators in Hilbert space are relevant to physics is of course very familiar to every particle theorist since the work of Wigner and Bargmann. That the formalism of operators in Hilbert space encompasses the variable geometries which underly gravity is the {\em leitmotiv} of our approach.

 In order to estimate the potential relevance of this approach to Quantum Gravity, one first needs to understand the physics underlying the problem of Quantum Gravity. There is an excellent article for this purpose: the paper \cite{woodard} explains how the problem arises when one tries to apply the perturbative method (which is so successful in quantum field theory) to the Lagrangian of gravity coupled with matter. Quoting from \cite{woodard}: ``Quantization of gravity is inevitable because part of the metric depends upon the other fields whose quantum nature has been well established".
 
   Two main points are that the presence of the other fields forces one, due to renormalization,  to add higher derivative terms of the metric to the Lagrangian and this in turns introduces at the quantum level an inherent instability  that would make the universe blow up. This instability is instantly fatal to an interacting quantum field theory. Moreover primordial inflation prevents one from fixing the problem by discretizing space at a very small length scale. What our approach permits is to develop a ``particle picture" for geometry; and a careful reading of the present paper should hopefully convince the reader that this particle picture stays very close to the inner workings of the Standard Model coupled to gravity. For now the picture is limited to the ``one-particle" description and there are deep purely mathematical reasons to develop the many particles picture. The main one is that the root of the one-particle picture, described by spectral triples, is $KO$-homology and the dual topological $KO$-theory (see \S \ref{notion of manifold}). The duality between the two theories is the origin of the quanta of geometry given by irreducible representations of the higher Heisenberg relation described in \S \ref{higherheis} below. As already mentioned in \cite{coinaugural}, algebraic $K$-theory, which is a vast refinement of the topological theory,  is begging for the development of a dual theory and one should expect profound relations between this dual theory and the  theory of interacting quanta of geometry. As a concrete point of departure, note that the deepest results on the topology of diffeomorphism groups of manifolds are given by the Waldhausen algebraic $K$-theory of spaces and we refer to \cite{DGM} for a unifying picture of algebraic $K$-theory. For this paper, we now we discuss in depth the problem of the co-existence of the discrete and the continuum in geometry. 

\vspace{0.1in}
{\bf{Acknowledgement.}}
 I am grateful to Joseph Kouneiher and Jeremy Butterfield  for their help in the elaboration of this paper.

\section{Prelude: the discrete and the continuum}
In this preliminary section we shall discuss two solutions  of the mathematical  problem of treating the continuous and the discrete in a unified manner. We first briefly present Grothendieck's solution: the notion of a topos which allowed him to treat in a unified manner ordinary topological spaces and the combinatorial structures arising in the world of arithmetic. We continue with a text of Grothendieck on Riemann as a prelude for  a re-reading of Riemann's inaugural lecture. We then explain how the quantum formalism provides another solution to the coexistence of discrete and continuous variables. 
\subsection{Grothendieck's solution: Topos}

 Grothendieck's solution to the problem of treating the continuous and the discrete in a unified manner is the notion of a Topos.  It does reconcile the usual idea of a topological space with that of a discrete combinatorial diagram. One does not concentrate on the space $X$ itself, with  its points etc... but rather on the ability of $X$ to define a variable set $Z_x$ depending on $x\in X$. When $X$ is an ordinary topological space such a ``variable set" indexed by $X$ is simply a sheaf of sets on $X$. But this continues to make sense starting from
 an abstract combinatorial diagram!  In short the key idea here is the idea of replacing $X$ by its role as a parameter space  
 
 \vspace{0.5cm}
 
 \centerline{``{\bf Space $X$ $\to$ Category of variable sets with parameter in $X$}" }
 
 \vspace{0.5cm}
 
 The abstract categories of such ``sets depending on parameters" fulfill almost all properties of the category of sets, except the axiom of the excluded middle, and encode in a faithful manner a topological space $X$ through the category of sheaves of sets on $X$. This new idea is amazing in its simplicity, its connection with logics and the richness of the new class of spaces that it uncovers.   In Grothendieck's own words (see ``R\' ecoltes et Semailles" \cite{RS,maclarty}) one can sense his amazement : 
{\em
\begin{quote}``Le ``principe nouveau'' qui restait \`a trouver, pour
consommer les \'epousailles promises par des f\'ees propices, ce n'\'etait autre aussi que ce ``lit''
spacieux qui manquait aux futurs \'epoux, sans que personne jusque-l\`a s'en soit seulement
aper\c cu$\ldots$

Ce ``lit \`a deux places'' est apparu (comme par un coup de baguette magique$\ldots$) avec
l'id\'ee du {\it topos}. Cette id\'ee englobe, dans une intuition topologique commune, aussi bien
les traditionnels espaces (topologiques), incarnant le monde de la grandeur continue, que
les (soi-disant) ``espaces'' (ou ``vari\'et\'es'') des g\'eom\`etres alg\'ebristes abstraits imp\'enitents,
ainsi que d'innombrables autres types de structures, qui jusque-l\`a avaient sembl\'e riv\'ees
irr\'em\'ediablement au ``monde arithm\'etique'' des agr\'egats ``discontinus'' ou ``discrets''.\end{quote}}

I would like to stress a key point of Grothendieck's idea of topos by using a metaphor. From his point of view, one understands a geometric space not by directly staring at it: no, the space remains at the back of the stage as a hidden schemer which governs the variability of every object at the front of the stage which is occupied by the usual suspects such as ``abelian groups" for instance. But once one studies these usual suspects in their new environment one finds that their fine properties reveal, from their relations with ordinary abelian groups,  the cohomology of the hidden parameter space.
Here the word ``ordinary" means ``independent of the parameter" and thus ordinary sets form part of the new set theory. This makes sense because a Grothendieck topos admits a unique morphism to the topos of sets. 

\subsection{Riemann}
In the prelude of ``R\' ecoltes et Semailles" \cite{RS}, Alexandre Grothendieck makes the following points on the search for relevant geometric models for physics and  on Riemann's lecture on the foundations of geometry: (see Appendix  \S \ref{app1} for the English translation)
{\em

\begin{quote}``Il doit y avoir d\'ej\`a quinze ou vingt ans, en feuilletant le modeste volume constituant l'\oe uvre compl\`ete
de Riemann, j'avais \'et\'e frapp\'e par une remarque de lui ``en passant''. Il y fait observer qu'il se pourrait
bien que la structure ultime de l'espace soit ``discr\`ete'', et que les repr\'esentations ``continues'' que nous
nous en faisons constituent peut-\^etre une simplification (excessive peut-\^etre, \`a la longue$\ldots$) d'une r\'ealit\'e plus complexe~; que pour l'esprit humain, ``le continu'' \'etait plus ais\'e \`a saisir que ``le discontinu'', et qu'il nous sert, par suite, comme une ``approximation'' pour appr\'ehender le discontinu. 

C'est l\`a une remarque d'une p\'en\'etration surprenante dans la bouche d'un math\'ematicien, \`a un moment o\`u le mod\`ele euclidien de l'espace physique n'avait jamais encore \'et\'e mis en cause~; au sens strictement logique, c'est plut\^ot le discontinu qui, traditionnellement, a servi comme mode d'approche technique vers le continu.

 Les d\'eveloppements en math\'ematique des derni\`eres d\'ecennies ont d'ailleurs montr\'e une symbiose bien plus intime entre structures continues et discontinues, qu'on ne l'imaginait encore dans la premi\`ere moiti\'e de ce si\`ecle. Toujours est-il que de trouver un mod\`ele ``satisfaisant'' (ou, au besoin, un ensemble de tels mod\`eles, se ``raccordant'' de fa\c con aussi satisfaisante que possible$\ldots$), que celui-ci soit ``continu'', ``discret'' ou de nature ``mixte'' -- un tel travail mettra en jeu s\^urement une grande imagination conceptuelle, et un flair consomm\'e pour appr\'ehender et mettre \`a jour des structures math\'ematiques de type nouveau. 

Ce genre d'imagination ou de ``flair'' me semble chose rare, non seulement parmi les physiciens (o\`u Einstein et Schr\"odinger semblent avoir \'et\'e parmi les rares exceptions), mais m\^eme parmi les math\'ematiciens (et l\`a je parle en pleine connaissance de cause).

Pour r\'esumer, je pr\'evois que le renouvellement attendu (s'il doit encore venir$\ldots$) viendra plut\^ot d'un math\'ematicien dans l'\^ame, bien inform\'e des grands probl\`emes de la physique, que d'un physicien. Mais surtout, il y faudra un homme ayant ``l'ouverture philosophique'' pour saisir le n\oe ud du probl\`eme. Celui-ci n'est nullement de nature technique, mais bien un probl\`eme fondamental de ``philosophie de la nature''.\end{quote}}

After reading the above text of Grothendieck, let us go to the relevant part of Riemann's Habilitation lecture on the foundations of geometry and explain why his great insight is, together with the advent of quantum mechanics, the best prelude  to the new paradigm of spectral triples, the basic geometric concept in NCG.

{\em

\begin{quote}  ``Wenn
aber eine solche Unabh\"{a}ngigkeit der K\"{o}rper vom Ort nicht
stattfindet, so kann man aus den Massverh\"{a}ltnissen im Grossen
nicht auf die im Unendlichkleinen schliessen; es kann dann in
jedem Punkte das Kr\"{u}mmungsmass in drei Richtungen einen
beliebigen Werth haben, wenn nur die ganze Kr\"{u}mmung jedes
messbaren Raumtheils nicht merklich von Null verschieden ist; noch
complicirtere Verh\"{a}ltnisse k\"{o}nnen eintreten, wenn die
vorausgesetzte Darstellbarkeit eines Linienelements durch die
Quadratwurzel aus einem Differentialausdruck zweiten Grades nicht
stattfindet. {\color{red} Nun scheinen aber die empirischen Begriffe, in
welchen die r\"{a}umlichen Massbestimmungen gegr\"{u}ndet sind,
der Begriff des festen K\"{o}rpers und des Lichtstrahls, im
Unendlichkleinen ihre G\"{u}ltigkeit zu verlieren; es ist also
sehr wohl denkbar, dass die Massverh\"{a}ltnisse des Raumes im
Unendlichkleinen den Voraussetzungen der Geometrie nicht
gem\"{a}ss sind, und dies w\"{u}rde man in der That annehmen
m\"{u}ssen, sobald sich dadurch die Erscheinungen auf einfachere
Weise erkl\"{a}ren liessen.}

Die Frage \"{u}ber die G\"{u}ltigkeit der Voraussetzungen der
Geometrie im Unendlichkleinen h\"{a}ngt zusammen mit der Frage
nach dem innern Grunde der Massverh\"{a}ltnisse des Raumes.  Bei
dieser Frage, welche wohl noch zur Lehre vom Raume gerechnet
werden darf, kommt die obige Bemerkung zur Anwendung, dass bei
einer discreten Mannigfaltigkeit das Princip der
Massverh\"{a}ltnisse schon in dem Begriffe dieser
Mannigfaltigkeit enthalten ist, bei einer stetigen aber anders
woher hinzukommen muss.  {\color{red} Es muss also entweder das dem Raume zu
Grunde liegende Wirkliche eine discrete Mannigfaltigkeit bilden,
oder der Grund der Massverh\"{a}ltnisse ausserhalb, in darauf
wirkenden bindenen Kr\"{a}ften, gesucht werden.}

Die Entscheidung dieser Fragen kann nur gefunden werden, indem
man von der bisherigen durch die Erfahrung bew\"{a}hrten
Auffassung der Erscheinungen, wozu \emph{Newton} den Grund
gelegt, ausgeht und diese durch Thatsachen, die sich aus ihr
nicht erkl\"{a}ren lassen, getrieben allm\"{a}hlich umarbeitet;
solche Untersuchungen, welche, wie die hier gef\"{u}hrte, von
allgemeinen Begriffen ausgehen, k\"{o}nnen nur dazu dienen, dass
diese Arbeit nicht durch die Beschr\"{a}nktheit der Begriffe
gehindert und der Fortschritt im Erkennen des Zusammenhangs der
Dinge nicht durch \"{u}berlieferte Vorurtheile gehemmt wird.

Es f\"{u}hrt dies hin\"{u}ber in das Gebiet einer andern
Wissenschaft, in das Gebiet der Physik, welches wohl die Natur
der heutigen Veranlassung nicht zu betreten erlaubt".\end{quote}}

This can be translated as follows:

``But if the
independence of bodies from position is not fulfilled, we cannot
draw conclusions from metric relations of the large, to those of
the infinitely small; in that case the curvature at each point
may have an arbitrary value in three directions, provided that
the total curvature of every measurable portion of space does not
differ sensibly from zero.  Still more complicated relations may
exist if we no longer assume that the line element is expressible as
the square root of a quadratic differential. {\color{red}  Now it seems that the
empirical notions on which the metric determinations of space
are based, the notion of  solid body and of  ray of light,
cease to be valid for the infinitely small.  We are therefore
quite free to assume that the metric relations of space in
the infinitely small do not comply with the hypotheses of
geometry; and we ought in fact to do this, if we can thereby
obtain a simpler explanation of phenomena.}

The question of the validity of the hypotheses of geometry in the
infinitely small is tied up with the question of the origin of
the metric relations of space.  In this last question, which we
may still regard as belonging to the doctrine of space, is found
the application of the remark made above; that in a discrete
manifold, the origin of its metric relations is given intrinsically, while in a continuous manifold, this origin
must come from outside.  {\color{red} Either therefore the reality which
underlies space must form a discrete manifold, or we must
seek the origin of its metric relations outside it, in the binding
forces which act upon it.}

The answer to these questions can only be obtained by starting from
the conception of phenomena which has hitherto been justified by
experiments, and which Newton assumed as a foundation, and by
making in this conception the successive changes required by
facts which it cannot explain.  Researches starting from general
notions, like the investigation we have just made, can only be
useful in preventing this work from being hampered by too narrow
views, and progress in knowledge of the interdependence of things
from being prevented by traditional prejudices.

This leads us into the domain of another science, of physics, into
which the object of this work does not allow us to enter today."


\subsection{The quantum and variability}\label{sect qvariables}

The originality of the quantum world (in which we actually live) as compared to its classical approximation, is already manifest at the experimental level by the ``imaginative 
randomness" of the results of experiments in the microscopic world. In order to appreciate this point consider the problem of manufacturing  a random number generator in such a way that even if an attacker happens to know the full details of the system the chance of reproducing the outcome is zero. This problem was solved concretely by  Bruno Sanguinetti, Anthony Martin, Hugo Zbinden, and Nicolas Gisin from the Group of Applied Physics, University of Geneva (see \cite{SMZG}). They invented : ``A generator of random numbers of quantum origin using technology compatible with consumer and portable electronics and whose  simplicity and performance  will make the widespread use of quantum random numbers a reality, with an important impact on information security."

\begin{figure}[H]
\begin{center}
\includegraphics[scale=1]{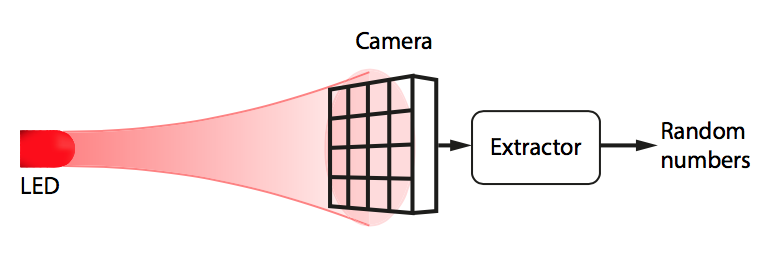}
\end{center}
\caption{The device uses the light emitted by a LED to produce a random number based on the quantum randomness of which cells of the camera are reached by emitted photons.  This figure is taken from \cite{SMZG}.\label{randomness} }
\end{figure}

This inherent randomness of the quantum world is not totally arbitrary since when the observable quantities that one measures happen to commute the usual classical intuition does apply. 
We owe to Werner Heisenberg the discovery\footnote{which he did while he was in the Island of Helgoland trying to recover from hay fever away from pollen sources} that the order of terms does matter when one deals with physical quantities which pertain to microscopic systems. We shall come back later in \S \ref{ncbonus} to the meaning of this fact but for now we retain that  
when manipulating the observables quantities for a microscopic system, the order of terms in a product plays a crucial role.  
\begin{figure}[H]
\begin{center}
\includegraphics[scale=0.2]{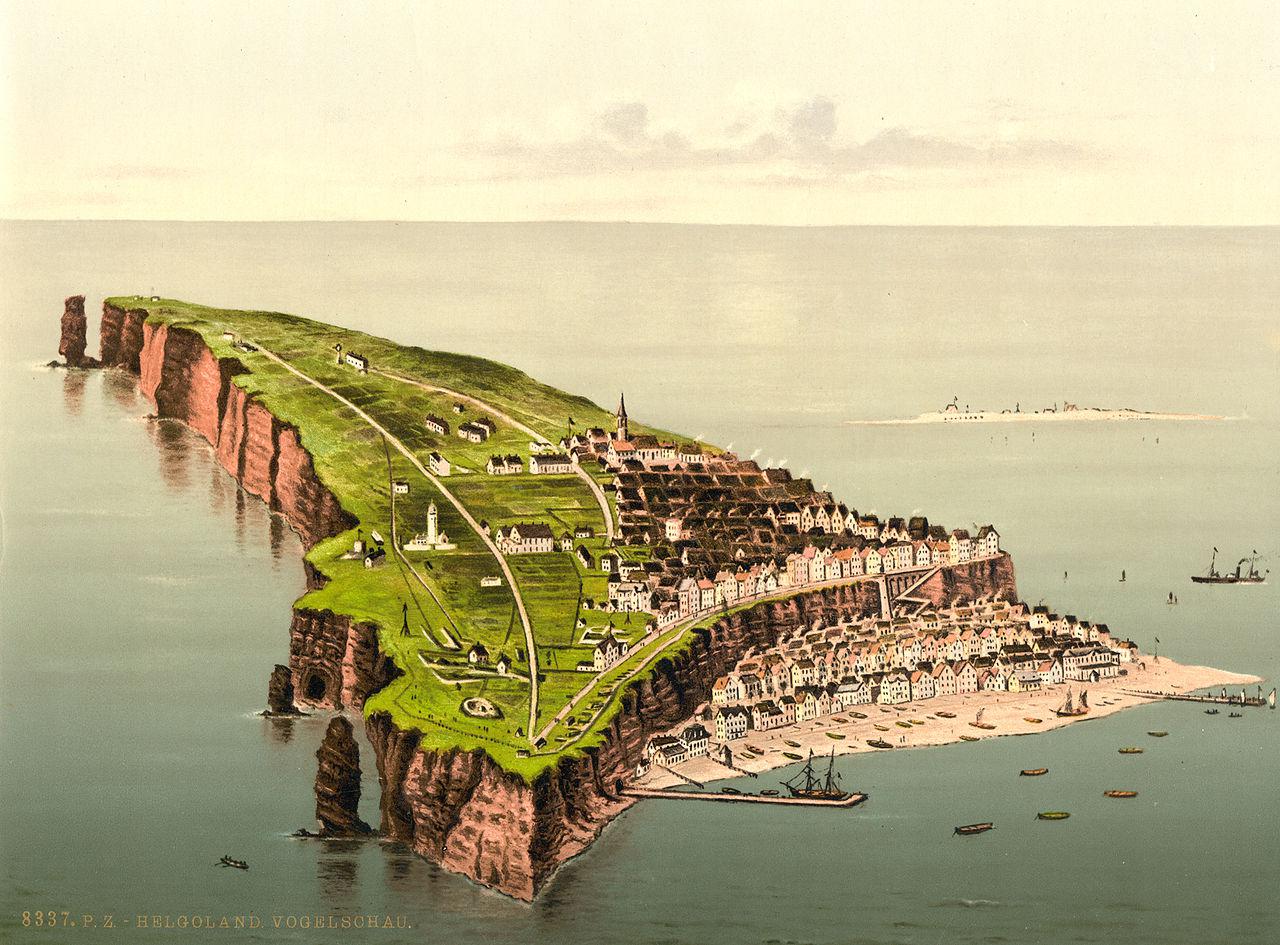}
\end{center}
\caption{Birdseye view, Helgoland, Germany, between 1890 and 1900. Image available from the United States Library of Congress's Prints and Photographs division, digital ID ppmsca.00573. \label{helgoland} }
\end{figure}
The commutativity of Cartesian coordinates does not hold in the algebra of coordinates on the phase space of a microscopic system. 
What Heisenberg discovered was that quantum observables obey the rules of matrix mechanics and  this led von Neumann to formalize quantum mechanics in terms of operators on Hilbert space. Let us explain now why this formalism actually provides a mathematical notion of ``real variable" which allows for the coexistence of continuous and discrete variables. Let us first display the defect of the classical notion.
In the classical formulation of real variables as maps from a set $X$ to the real numbers $\R$, the set $X$ has to be uncountable if some variable has continuous range. But then for any other variable with countable range some of the multiplicities are infinite. This means that discrete and continuous variables cannot coexist in this classical formalism. 

Fortunately everything is fine and this problem of treating continuous and discrete variables on the same footing is completely solved using the formalism of quantum mechanics which  provides another solution and treats directly the notion of real variable. The key replacement is 
\vspace{0.5cm}
 
 \centerline{``{\bf Real Variable $\to$ Self Adjoint Operator in Hilbert space}" }
 
 \vspace{0.5cm}

All the usual attributes of real variables such as their range, the number of times a real number is reached as a value of the variable etc... have a perfect analogue in the quantum mechanical setting.
The range is the spectrum of the operator, and the spectral multiplicity gives the number of times a real number is reached. It is very comforting for instance that one can compose any measurable (Borel) map $h:\R\to \R$ with any self-adjoint operator $H$ so that $h(H)$ makes sense and has the expected property of the composed real variable. In the early times of quantum mechanics, physicists had a clear intuition of this analogy between operators in Hilbert space (which they called q-numbers) and variables.
 Note that the choice of Hilbert space is irrelevant here since all separable infinite dimensional Hilbert spaces are isomorphic.  
 \bigskip 
 
\begin{center}
\begin{tabular}{|c|c|}
\hline &  \\
{\bf \color{blue}Classical } & {\bf \color{blue}Quantum }
\\
&\\
\hline &  \\
 Real variable & Self-adjoint  \\
$f:X\to \R$ & operator in Hilbert space\\
&\\
\hline &  \\
 Possible values &  Spectrum of   \\
 of the variable & the operator  \\
&\\ \hline &  \\
 Algebraic operations  & Algebra of operators \\
on functions & in Hilbert space\\
&\\ \hline 
\end{tabular}
\end{center}

\bigskip 

In fact it is the uniqueness of the separable infinite dimensional Hilbert space that cures the above problem of coexistence of discrete and continuous variables: $L^2[0,1]$ is the same as $\ell^2(\N)$, and variables with continuous range (such as the operator of multiplication by $x\in [0,1]$) coexist happily with variables with countable range (such as the operator of multiplication by $1/n, n\in \N$),  but they do not commute!

It is only because one drops commutativity that variables with continuous range can coexist with variables with countable range. 
 The only new fact is that they do not commute, and the real subtlety is in their algebraic relations.

 What is surprising is that the new set-up immediately provides a natural home for the ``infinitesimal variables": and here the distinction between ``variables" and numbers (in many ways this is where the point of view of Newton is more efficient than that of Leibniz) is essential.
 It is worth quoting Newton's definition of variables and of infinitesimals, as opposed to Leibniz:
{\em

\begin{quote}``In a certain problem, a variable is the quantity that takes an infinite number of values which are quite determined by this problem and are arranged in a definite order"\end{quote}

\begin{quote} ``A variable is called infinitesimal if among its particular values one can be found such that this value itself and all following it are smaller in absolute value than an arbitrary given number"\end{quote}}

Indeed it is perfectly possible for an operator to be ``smaller than epsilon for any epsilon" without being zero. This happens when the norm of the restriction of the operator to subspaces of finite codimension tends to zero when these subspaces decrease (under the natural filtration by inclusion). The corresponding operators are called ``compact" and they share with naive infinitesimals all the expected algebraic properties. \bigskip
\begin{center}
\begin{tabular}{|c|c|}
\hline &  \\
{\bf \color{blue}Classical } & {\bf \color{blue}Quantum }
\\
&\\
\hline &  \\
 Infinitesimal  & Compact  \\
variable & operator in Hilbert space\\
&\\
\hline &  \\
 Infinitesimal of &  $\mu_n(T)$ of size $n^{-\alpha}$   \\
 order $\alpha$ & when $n \to \infty$  \\
&\\ \hline &  \\
  Integral of  & $\displaystyle{\int\!\!\!\!\!\! -} T =$ coefficient of \\
function $\int f(x)dx$ & $\log(\Lambda)$ in $\Tr_\Lambda$(T)\\
&\\ \hline 
\end{tabular}

\end{center}

\bigskip
Indeed they form a two-sided ideal of the algebra of bounded operators in Hilbert space and the only property of the naive infinitesimal calculus that needs to be dropped is the commutativity. 

The calculus of infinitesimals fits perfectly into the operator formalism of quantum mechanics where compact operators play the role of infinitesimals, with order governed by the rate of decay of the characteristic values,  and where the logarithmic divergences familiar in physics give the substitute for integration of infinitesimals of order one, in the form of the Dixmier trace and Wodzicki's residue. We refer to \cite{Co-book} for a detailed description of the new integral $\displaystyle{\int\!\!\!\!\!\! -}$.

\section{The spectral paradigm}


Before we start the ``inward bound" trip \cite{Pais} to very small distances, it is worth explaining how the spectral point of view  helps  also when dealing with issues connected to large astronomical distances.

\begin{figure}[H]
\begin{center}
\includegraphics[scale=0.8]{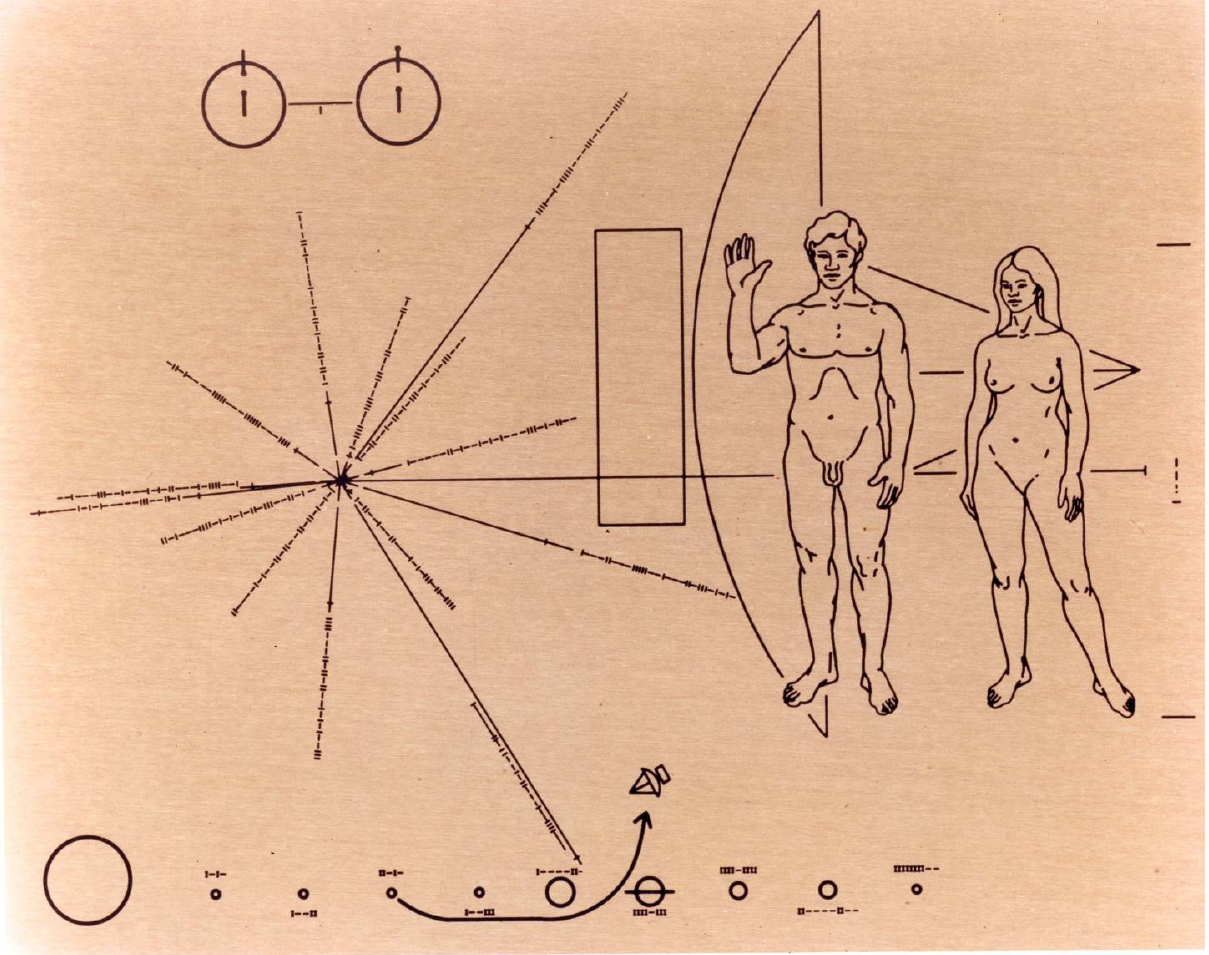}
\end{center}
\caption{The pioneer 4 probe, picture from NASA : 668774 main pioneer plaque. \label{pioneer} }
\end{figure}

The simple question ``Where are we?" does not have such a simple answer since giving our coordinates in a specific chart is not an invariant manner of describing our position. We refer to Figure \ref{pioneer} for one attempt at an approximate  answer.

In fact it is not obvious how to solve two mathematical questions which naturally arise in this context:
\begin{enumerate}
\item Can one specify a geometric space in an invariant manner?
\item Can one specify a point of a geometric space in an invariant manner?	
\end{enumerate}

\subsection{Why Spectral}

\begin{figure}[H]
\centering
\begin{minipage}{.5\textwidth}
  \centering
  \includegraphics[width=.66\linewidth]{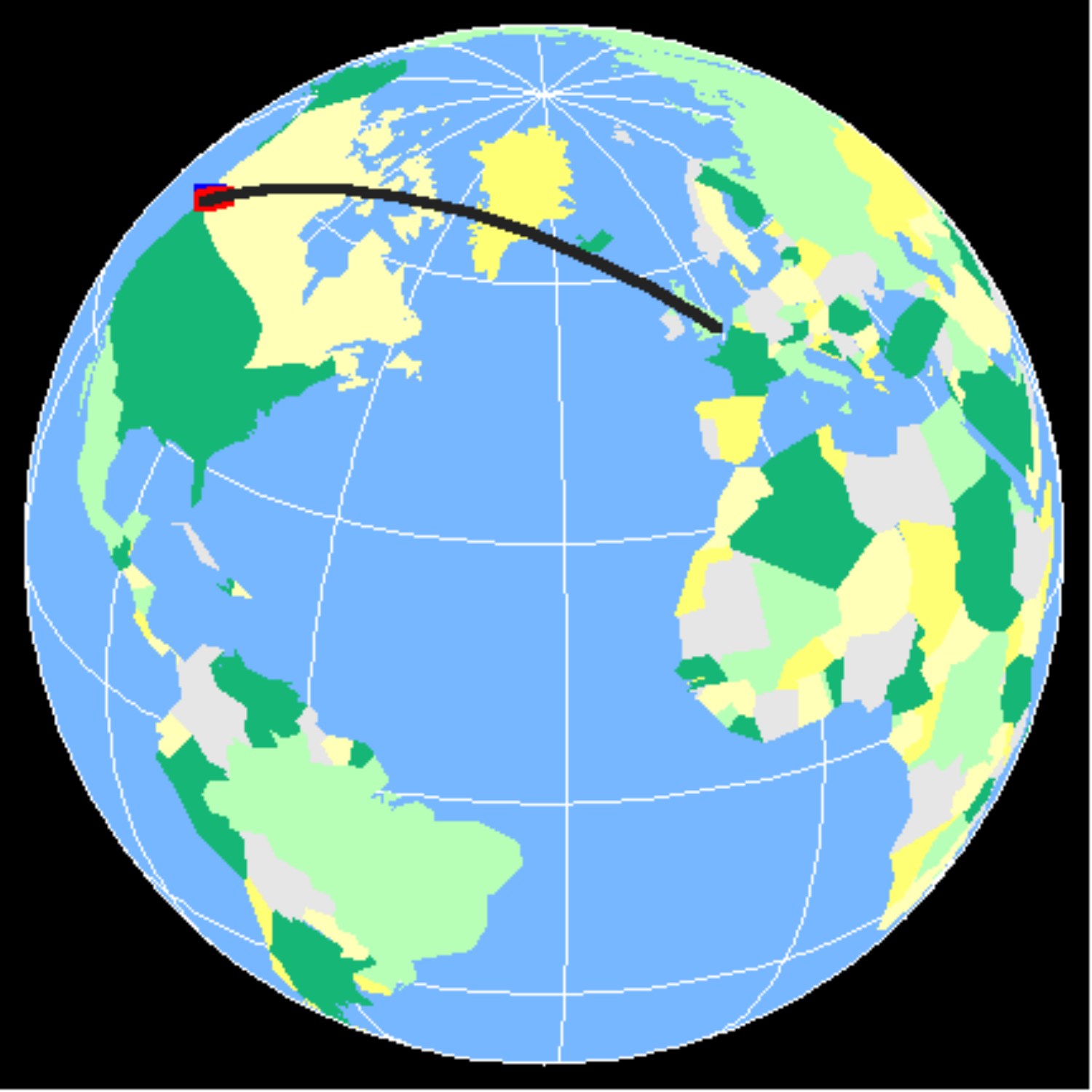}
  \captionof{figure}{Geodesic}
  \label{geodesic}
\end{minipage}%
\begin{minipage}{.5\textwidth}
  \centering
  \includegraphics[width=.64\linewidth]{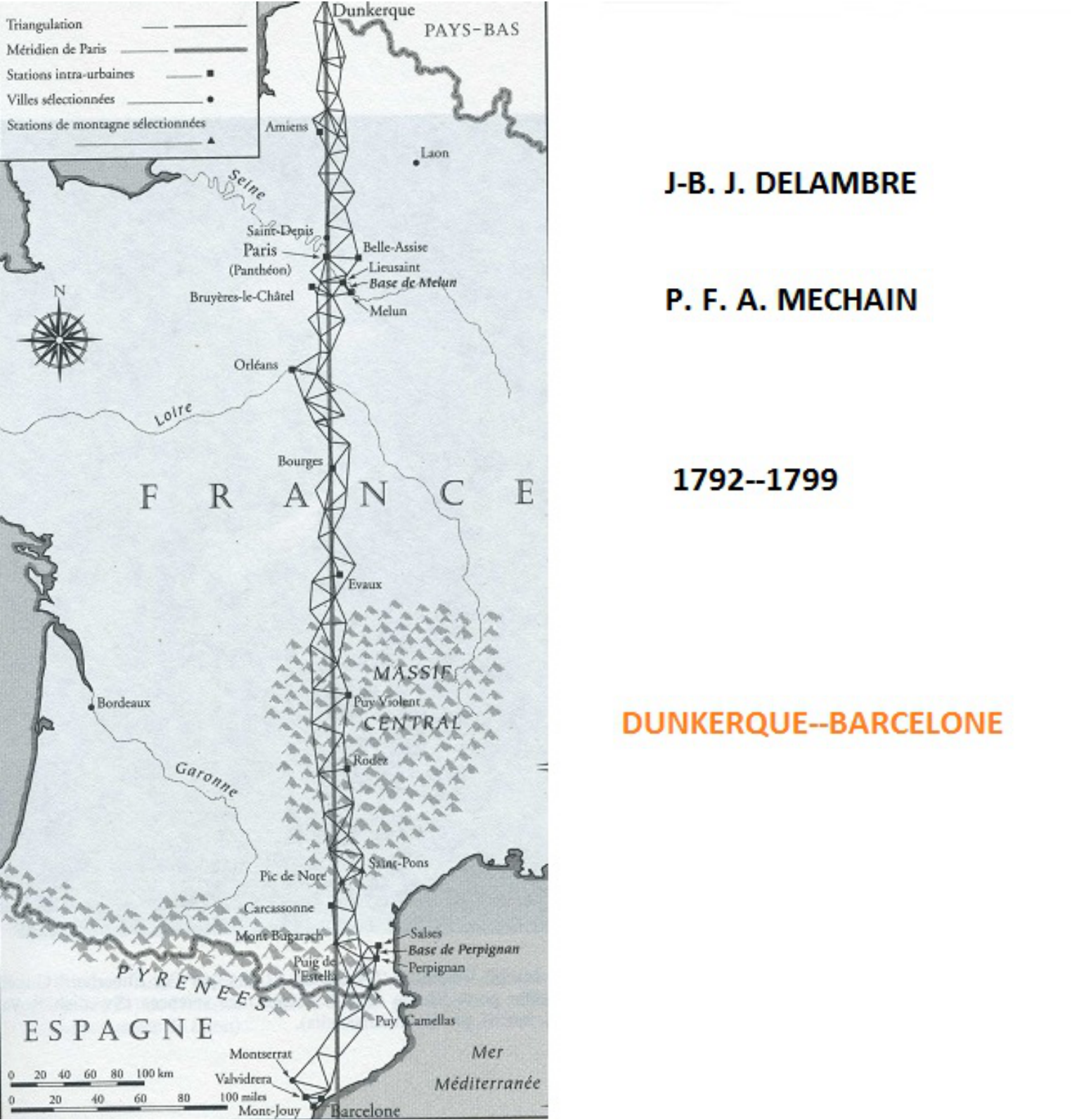}
  \captionof{figure}{Delambre and Mechain}
  \label{geodesic1}
\end{minipage}
\end{figure}

Given a compact Riemannian space one obtains a slew of geometric invariants of the space by considering the spectrum of natural operators such as the Laplacian. 
The obtained list of numbers is a bit like a scale associated to the space  as made clear by Mark Kac in his famous paper\footnote{Kac, Mark (1966), "Can one hear the shape of a drum?", American Mathematical Monthly 73 (4, part 2): 1--23} ``Can one hear the shape of a drum?".
It is well known however since a famous one page paper\footnote{Milnor, John (1964), "Eigenvalues of the Laplace operator on certain manifolds", Proceedings of the National Academy of Sciences of the United States of America 51} of John Milnor that the spectrum
of operators, such as the Laplacian, does not suffice to characterize a compact
Riemannian space. But it turns out that the missing information is encoded by
the relative position of two abelian algebras of operators in Hilbert space.
Due to a theorem of von Neumann, the algebra of multiplication by all measurable
bounded functions acts in Hilbert space in a unique manner, independent of the
geometry one starts with. Its relative position with respect to the other
abelian algebra given by all functions of the Laplacian suffices to recover the
full geometry, provided one knows the spectrum of the Laplacian. For some
reason which has to do with the inverse problem, it is better to work with the
Dirac operator; and as we shall explain now, this gives a guess for a new incarnation of the ``line element". The Riemannian paradigm is based on the Taylor expansion in local coordinates
of the square of the line element, and in order to measure the distance between
two points one minimizes the length of a path joining the two points as in Figure \ref{geodesic}
\begin{equation}\label{riemann distance}
d(a,b)=\,{\rm Inf}\,\int_\gamma\,\sqrt{g_{\mu\,\nu}\,dx^\mu\,dx^\nu}	
\end{equation}

Great efforts were done at the time of the French revolution in order to obtain a 
sensible unification of the various units of length that were in use across the country. It was decided (by Louis XVI, under the advice of Lavoisier) to take, as a unit, the length $L$ such that $4\times 10^6 L$ would be the circumference of the earth. After using as a preliminary reduction the computation of angles from astronomical observations to reduce the actual measurement to a smaller portion of meridian, a team was sent out in 1792 to make the precise measurement of the distance between Dunkerque in the north of France and Barcelona in Spain; see Figure \ref{geodesic1}. This measurement\footnote{I refer the reader  to \cite{meter} for a very interesting and more detailed account of the story of the measurement performed by Delambre and M\' echain}
 resulted in an incarnation of $L$ as a concrete platinum bar that was kept in Pavillon de Breuteuil near Paris. I remember learning in school this definition of the ``meter".  

However it turned out that in the $1930$'s,  physicists were able to decide that the above choice of $L$ was no good. Not only because it would seem totally unpractical if we would for instance try to transmit its definition to a far distant star, but for a more pragmatic reason, they observed that the concrete platinum bar defining $L$  actually had a non-constant length! This observation was done by comparing it with a specific wave length of Krypton. 
\begin{figure}[H]
\begin{center}
\includegraphics[scale=1]{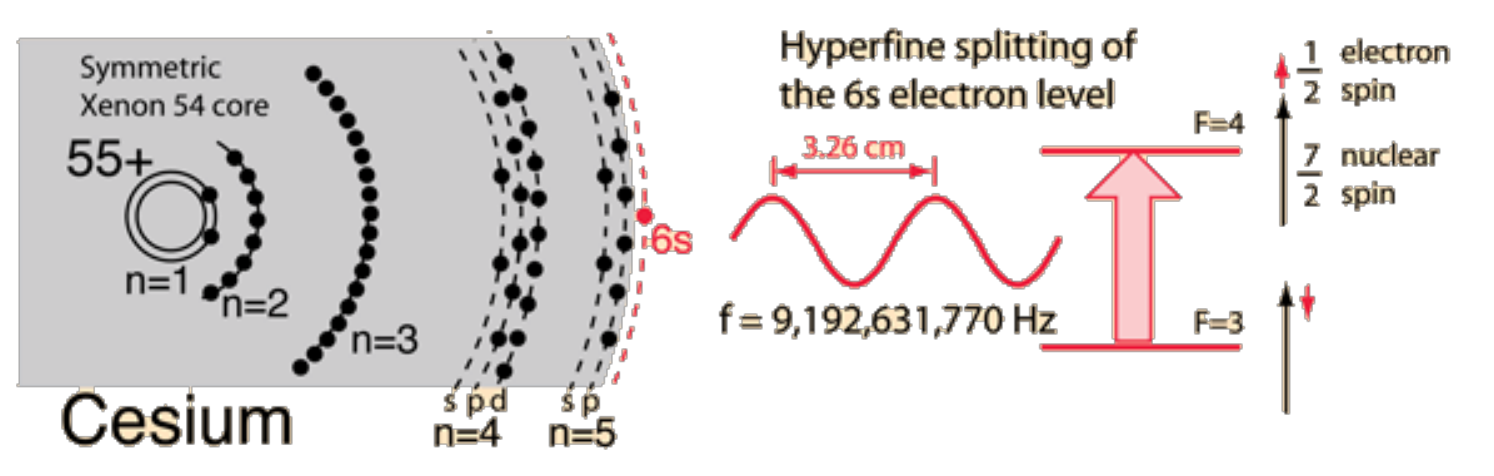}
\end{center}
\caption{Meter $\to$ Wave length, the 13th CGPM (1967)
uses hyperfine levels of Cesium (C133). Adaptation of original found at \url{hyperphysics.phy-astr.gsu.edu}\label{unit spec} }
\end{figure}

Then it took some time until they decided to take the obvious step: to replace $L$ by the wavelength of a specific atomic transition (the chosen one is called 2S$_{1/2}$ of Cesium 133), as was done in 1967. 

More precisely, this hyperfine transition  is used to define the second as the duration of 9 192 631 770 periods of the radiation corresponding to the transition between the two hyperfine levels of the ground state of the cesium 133 atom. 

Moreover the speed of light is set to the value of $299792458$ meters/second, which thus defines the meter as the length of the path travelled by light in vacuum during a time interval of 1/299 792 458 of a second, \ie the meter is
$$
\frac{9 192 631 770}{299792458}=\frac{656616555}{21413747}\sim 30.6633 
$$
times the wavelength of the hyperfine transition of Cesium (which is of the order of $3.26 $cm).

What is manifest with this new choice of $L$ is that one now has a chance to be able to communicate our ``unit of length" with aliens without telling them to come to Paris etc... Probably in fact this issue should motivate us to choose a chemical element such as hydrogen which is far more common in the universe than Cesium.
One striking advantage of the new choice of $L$ is that it is no longer ``localized" (as it was before near Paris) and is available anywhere using the constancy of the spectral properties of atoms. It will serve us as a motivation for our spectral paradigm. 

\subsection{The line element}

The presence of the square root in \eqref{riemann distance} is the witness of Riemann's prescription for the square of the line element as $ds^2=g_{\mu\,\nu}\,dx^\mu\,dx^\nu$. In the spectral framework the extraction of the square root of the Laplacian goes back to Hamilton who already wrote, using his quaternions, the key combination $$D=i \partial_x+ j \partial_y+ k \partial_z$$ 
The conceptual algebraic device for extracting the square root of sums of squares such as $X^2+Y^2$ is provided by the Clifford algebra where the anti-commutation
$XY=-YX$ provides the simplification $(X+Y)^2=  X^2+Y^2$.

\begin{figure}[H]
\begin{center}
\includegraphics[scale=1.3]{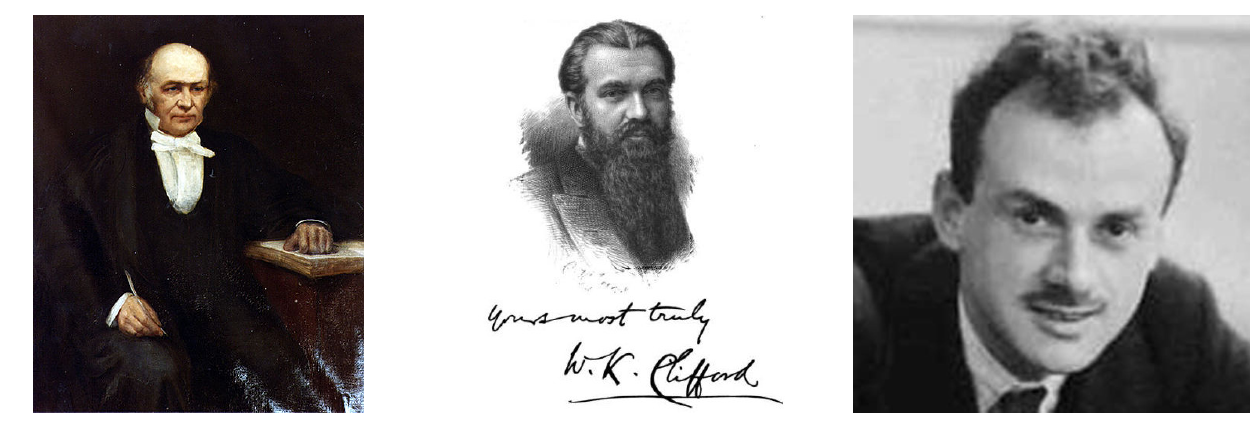}
\end{center}
\caption{Hamilton \cite{hankins}, Clifford, Dirac \label{dirac} }
\end{figure}

P. Dirac showed how to extract the square root of the Laplacian in order to obtain a relativistic form of the Schr\"odinger equation. For curved spaces Atiyah and Singer devised a general formula for the Dirac operator on a spin Riemannian manifold and this provides us with our prescription: the line element is the propagator
$$
ds=D^{-1}
$$
(where one takes the value $0$ on the kernel). This allows us to measure distances and \eqref{riemann distance} becomes
\begin{equation}\label{dirac distance}
d(a,b)=\,{\rm Sup}\,\vert f(a)-f(b)\vert, f \ \text{such} \; \text{that}\ \Vert [D,f]\Vert\leq 1.
\end{equation}
which gives the same answer as \eqref{riemann distance} and is 
 a ``Kantorovich dual" of the usual formula. But we now have the possibility to define and measure distances without the need of paths joining two points as in \eqref{riemann distance}. And indeed one finds plenty of examples of totally disconnected spaces in which the new formula \eqref{dirac distance} makes sense and gives sensible results while \eqref{riemann distance} would not, due to the absence of connected arcs. 
\begin{figure}[H]
\begin{center}
\includegraphics[scale=0.4]{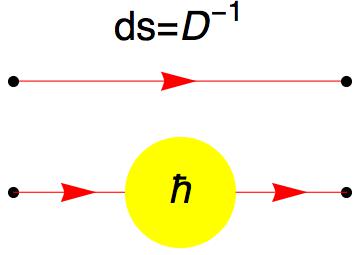}
\end{center}
\caption{Line Element \label{line} }
\end{figure}
The link of this new definition of distances (and hence of geometry) with the quantum world appears in many ways: first the ``line element" ought to be an ``infinitesimal". This indeed fits  since in a compact Riemannian spin manifold the above operator $ds$  is compact \ie infinitesimal as explained in \S \ref{sect qvariables}. But there are two more facts which help us to appreciate the relevance of the new concept: both are displayed in Figure \ref{line}. In the upper part the directed line is a common ingredient of Feynman diagrams, it represents the internal legs of fermionic diagrams and is called the ``fermion propagator". Physically it represents a very tiny interval in which the interaction takes place. Mathematically it is our ``ds" (modulo a bit of agility in understanding the physics language  and in particular the need to pass from the Minkowski signature to the Euclidean one). The lower part of Figure \ref{line} displays an even more important feature: the above fermionic propagator undergoes quantum corrections due to its role in quantum field theory and we can interpret these corrections as quantum corrections to the geometry!

\subsection{The  bonus from non-commutativity}\label{ncbonus}

In algebra the commutativity assumption often appears as a welcome simplification 
which makes many algebraic manipulations much easier.  But in fact we should realize that  our use of the written language makes us perfectly familiar with non-commutativity. The advantage, as far as meaning is concerned, of paying attention to the order of terms, becomes clear when considering anagrams \ie writings which become equal when ``abelianized" but nevertheless have quite different meanings when the order of terms is respected. Here is a recent anagram which can be found in ``Anagrammes pour lire dans les pens\' ees" by Raphael Enthoven and Jacques Perry-Salkow,

\vspace{0.8cm}

{\color{blue}

\centerline{\it ``ondes gravitationnelles"}

\vspace{0.4cm}

\centerline {\it``le vent d'orages lointains"}}

\vspace{0.8cm}

When we permit ourselves  to commute the various letters involved in each of these phrases we find the same result:
$$
a^2 d e^3 g i^2 l^2 n^3 o^2 r s^2 t^2 v
$$
This shows that in projecting a phrase in the commutative world one looses an enormous amount of information encoded by non-commutativity. 
Natural languages respect non-commutativity and a phrase is a much more informative datum than its commutative algebraic shadow. 

Here are two more key features of the noncommutative world:
\begin{enumerate}
\item Non-commuting discrete variables of the simplest kind generate continuous variables.
\item A noncommutative algebra possesses inner automorphisms. 	
\end{enumerate}
We always think of variables through their representations as operators in Hilbert space as explained  in \S \ref{sect qvariables} and since the product of two self-adjoint operators is not self-adjoint unless they commute, one deals with algebras $\cA$ which are $*$-algebras \ie which  are endowed with an antilinear involution which obeys the rule $(xy)^*=y^*x^*$ for any $x,y\in\cA$. The simplest noncommutative algebra of this kind is $M_2(\C)$ the algebra of $2\times 2$ matrices 
$$
a=\left(
\begin{array}{cc}
 a_{11} & a_{12} \\
 a_{21} & a_{22} \\
\end{array}
\right),  \ b=\left(
\begin{array}{cc}
 b_{11} & b_{12} \\
 b_{21} & b_{22} \\
\end{array}
\right), \ ab=\left(
\begin{array}{cc}
 a_{11} b_{11}+a_{12} b_{21} & a_{11} b_{12}+a_{12} b_{22} \\
 a_{21} b_{11}+a_{22} b_{21} & a_{21} b_{12}+a_{22} b_{22} \\
\end{array}
\right)
$$
and the antilinear involution is given using the complex conjugation $z\mapsto \bar z$ by the conjugate transpose, \ie 
$$
a^*=\left(
\begin{array}{cc}
  \bar a_{11} & \bar a_{21} \\
 \bar a_{12} &\bar a_{22} \\
\end{array}
\right)
$$
This algebra $M_2(\C)$ only represents discrete variables taking at most two values but as soon as one adjoins another non-commuting variable  $Y$, such that $Y=Y^*$ and $Y^2=1$ one generates all matrix valued functions on the two-sphere.

To be more precise, write the above generic matrix in the form $a= a_{11} e_{11}+a_{12} e_{12}+a_{21} e_{21}+a_{22} e_{22}$ where the $e_{ij}\in M_2(\C)$, and the coefficients $a_{ij}\in\C$ are complex numbers. Then using algebra one can write  $Y= y_{11} e_{11}+y_{12} e_{12}+y_{21} e_{21}+y_{22} e_{22}$ where the $y_{ij}$ are no longer complex numbers but commute with $M_2(\C)$. For instance $y_{11}=e_{11} Y e_{11}+e_{21} Y e_{12}$.  One imposes the additional condition that the trace of $Y$ is zero, \ie that $y_{11}+y_{22}=0$. It is then an exercise using the relations $Y=Y^*$ and $Y^2=1$, to show that the $C^*$-algebra generated by the $y_{ij}$ is the algebra $C(S^2)$ of continuous functions on the two sphere $S^2$. It contains of course plenty of ``continuous variables" and the traditional sup norm of complex valued functions is 
$$
{\rm Sup}_{x\in S^2} \vert f(x)\vert = {\rm Sup}_\pi \Vert \pi(f)\Vert 
$$
where in the right hand side $\pi$ runs through all Hilbert space representations (compatible with the involution $*$) of the above relations. One obtains all continuous functions by completion and thus one keeps inside the algebra $C(S^2)$ the nicer smooth functions such as those algebraically obtained from the $y_{ij}$. The sphere itself is recovered as the {\bf Spectrum} of the algebra, and the points of the sphere are the characters \ie the morphisms of involutive algebras to $\C$. 

This is a prototype example of how a connected space (here the two sphere $S^2$) can spring out of the discrete (here $M_2(\C)$ and the two valued variable $Y$) due to non-commutativity. Note also the compatibility of the two notions of spectrum. Indeed for $f$ in the commutative algebra generated by the $y_{ij}$, the spectrum of the operator $\pi(f)$ is the image by the corresponding function on $S^2$ of the support of the representation $\pi$ which is a closed subset of the spectrum of the algebra.
To put this in a suggestive manner: what happens is that the geometric space $S^2$ appeared in a spectral manner and from familiar players of the quantum world: the algebra $M_2(\C)$, for instance, is familiar from spin systems.

There is another great bonus from non-commutativity: the natural algebra which springs out of the non-commuting $M_2(\C)$ and $Y$ discussed above is not the algebra generated by the $y_{ij}$ but the algebra generated by $M_2(\C)$ and $Y$. It contains the former but is larger and gives the algebra $C(S^2,M_2(\C))$  of matrix valued continuous functions on the two sphere. If we take the subalgebra of smooth functions $\cA=C^\infty(S^2,M_2(\C))$   (which is canonically obtained inside $C(S^2,M_2(\C))$ by applying the smooth functional calculus to the generators)  and one looks at its automorphism group\footnote{Compatible with the $*$-operation.}, one finds that it fits in an exact sequence 
$$
1\to \Int(\cA)\to \Aut(\cA) \to \Out(\cA)\to 1.
$$
Such an exact sequence exists for any non-commutative $*$-algebra, the inner automorphisms $\Int(\cA)$ are those of the form $x\mapsto uxu^*$ where $u\in \cA$ is a unitary element \ie fulfills $uu^*=u^*u=1$. The nice general fact is that these automorphisms always form a normal subgroup of the group $\Aut(\cA)$ and the quotient group  $\Out(\cA)$ is called the group of outer automorphisms of $\cA$. Now when one computes these groups in our example \ie for $\cA=C^\infty(S^2,M_2(\C))$, one finds that  the group $\Out(\cA)$ is the group of diffeomorphisms ${\rm Diff}(S^2)$ while $\Int(\cA)$ is the group of smooth maps from $S^2$ to the Lie group $PSU(2)$ whose Lie algebra is $su(2)$. Thus we witness in this example the marriage of the gauge group of gravity \ie the diffeomorphism group, with the gauge group of matter \ie here of an $su(2)$-gauge theory.   
\subsection{The notion of manifold}\label{notion of manifold}\label{notionof manifold}
The notion of spectral geometry has deep roots in pure mathematics. They have to do with the  understanding of the notion of (smooth) manifold. While this notion is simple to define in terms of local charts \ie by glueing together open pieces of finite dimensional vector spaces, it is much more difficult and instructive to arrive at a global understanding. To be specific we now discuss the notion of a compact oriented smooth manifold.

 What one does is to detect global properties of the underlying space with the goal of characterizing manifolds. At first one only looks at the space up to homotopy. The broader category of ``manifolds" that one first obtains is that of ``Poincar\' e complexes" \ie of CW complexes $X$ which satisfy Poincar\' e duality with respect to the fundamental homology class  with coefficients in $\Z$. It is important to take into account the fundamental group $\pi_1(X)$, and to assume Poincar\' e duality with arbitrary local coefficients. In the simply connected case, a result of Spivak \cite{spivak} shows the existence (and  uniqueness up to stable fiber homotopy equivalence) of a spherical fibration, called the Spivak normal bundle $p:E\to X$. Such a fibration satisfies the covering homotopy property and each fiber $p^{-1}(x)$ has the homotopy type of a sphere. At this point one is still very far from dealing with a manifold and the obstruction to obtain a smooth manifold in the given homotopy type is roughly the same as that of finding a vector bundle whose associated spherical fibration is $p:E\to X$. This follows from the work of Novikov and Browder at the beginning of the 1960's.  There are important nuances between piecewise linear (PL) and smooth but they do not affect the 4-dimensional case in which we are interested. 
 
 The first key root of the notion of ``spectral geometry" is a result of D. Sullivan (see \cite{MS}, epilogue) that a PL-bundle is the same thing (modulo the usual ``small-print" qualifications at the prime $2$, \cite{Siegel}) as a spherical fibration together with a $KO$-orientation. What we retain is that the key property of a ``manifold" is not Poincar\' e duality in ordinary homology but is Poincar\' e duality in the finer theory called $KO$-homology. To understand how much finer that theory is, it is enough to state that the fundamental class $[X]\in KO_*(X)$ contains all the information about the Pontrjagin classes of the manifold and these are not at all determined by its homotopy type: in the simply connected case only the signature class is fixed by the homotopy type.
 
  Here comes now the second crucial root of the notion of spectral geometry from pure mathematics. In their work on the index theorem, Atiyah and Singer understood that operators in Hilbert space  provide the right realization for $KO$-homology cycles \cite{Atiyah, Singer}. Their original idea was developed by Brown-Douglas-Fillmore, Voiculescu, Mischenko and 
acquired its definitive form in the work of Kasparov at the end of the 1970's. The great new tool is  bivariant Kasparov theory, but as far as  $K$-homology cycles are concerned\footnote{The nuance between $K$ and $KO$ is important and gives rise to the real structure discussed in the next section} the right notion is already in Atiyah's paper \cite{Atiyah}: A $K$-homology cycle on a compact space $X$ is given by a representation of the algebra $C(X)$ (of continuous functions on $X$) in a Hilbert space $\cH$, together with a Fredholm operator $F$ acting in the same Hilbert space fulfilling some simple compatibility condition (of commutation modulo compact operators) with the action of $C(X)$. One striking feature of this representation of $K$-homology cycles is that the definition does not make any use of the commutativity of the algebra $C(X)$.

 At the beginning of the 1980's, motivated by numerous examples of noncommutative spaces arising naturally in geometry from foliations or in physics from the Brillouin zone in the work of Bellissard on the quantum Hall effect, I realized that specifying an unbounded representative of the Fredholm operator gave the right framework for spectral geometry. The corresponding $K$-homology cycle only retains the stable information and is insensitive  to deformations while the unbounded representative encodes the metric aspect. These are the deep mathematical reasons which are the roots of the notion of spectral triple. 
 \subsection{Real structure}\label{sectreal}
The
additional structure on a $K$-homology cycle that upgrades it into a $KO$-homology cycle  is given by requiring a {\em real structure} \cite{Coreal},
\ie an antilinear unitary operator $J$ acting in $\cH$ which plays the same
role and has the same algebraic properties  as the charge conjugation operator
in physics.
\begin{itemize}

\item In physics $J$ is the charge conjugation operator.

\item It is deeply related  to Tomita's operator which conjugates the algebra
with its commutant. The basic relation always satisfied is Tomita's relation:
$$
[a,b^{\rm op}]=0\qqq a,  b \in\mathcal{A}, b^{\rm op}:= Jb^*J^{-1}.
$$
\item In $KO$-homology, one obtains a $KO$-homology cycle for the algebra 
$\cA\otimes \cA^{\rm op}$
and an intersection form:
$$
K(\cA)\otimes K(\cA)\to \Z, \  {\rm Index}(D_{e\otimes f})
$$
\end{itemize}
\begin{figure}[H]
\begin{center}
\includegraphics[scale=0.6]{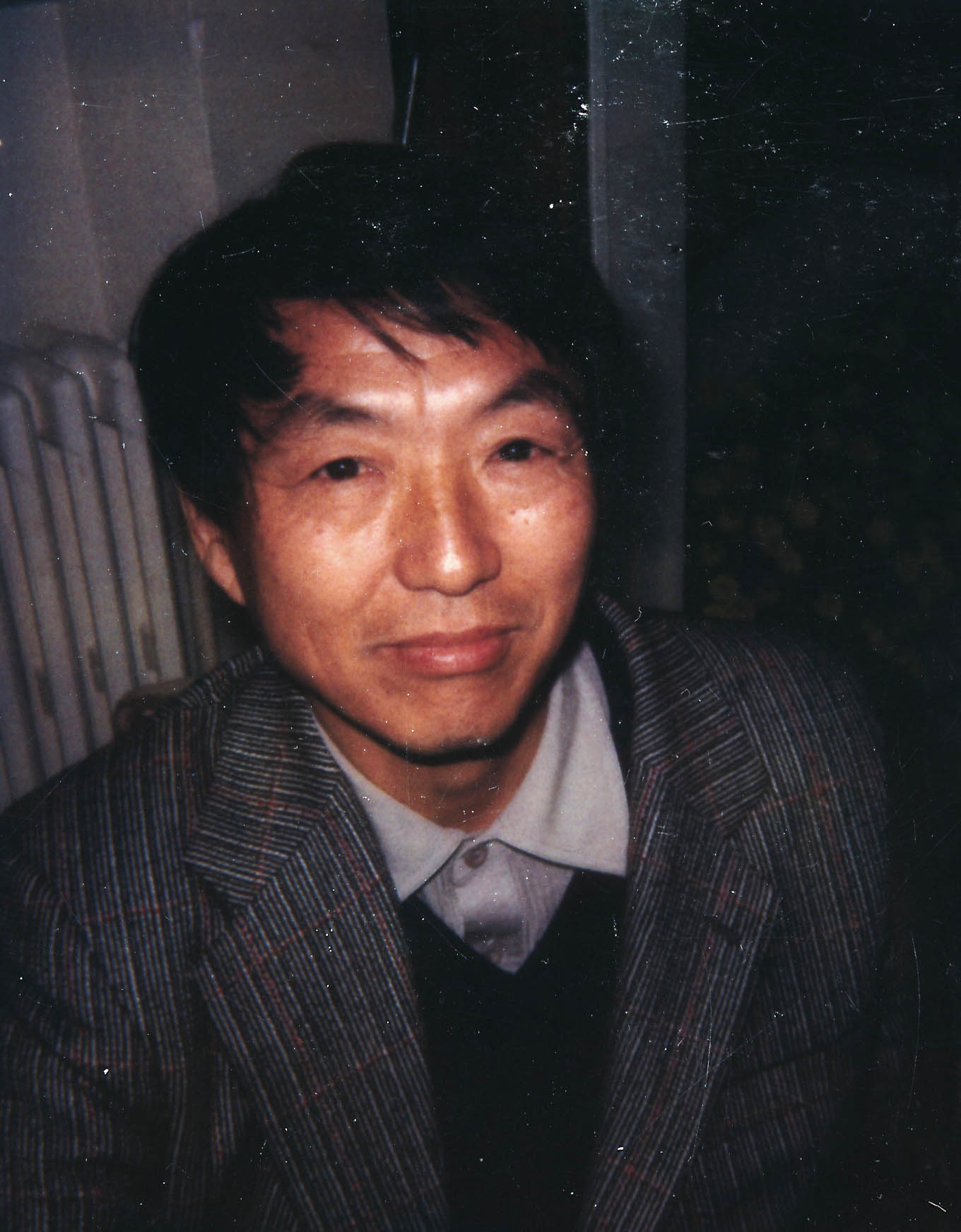}
\end{center}
\caption{Minoru Tomita \label{tomita} }
\end{figure}

In the even case, the chirality operator $\gamma$ plays an important role, both $\gamma$ and $J$ are decorations of the spectral triple.

 The following further relations hold  for $D,J$ and $\gamma$
$$
J^2=\varepsilon\,, \ DJ=\varepsilon^{\prime}JD,\quad
J\,\gamma=\varepsilon^{\prime\prime }\gamma J,\quad D\gamma=-\gamma D
$$

 The values of the three signs $\varepsilon,\varepsilon^{\prime
},\varepsilon^{\prime\prime}$ depend only, in the classical case of spin
manifolds, upon the value of the dimension $n$ modulo $8$ and are given in the
following table:
\begin{center}
 \begin{tabular}
[c]{|c|rrrrrrrr|}\hline \textbf{n } & 0 & 1 & 2 & 3 & 4 & 5 & 6 &
7\\\hline\hline
$\varepsilon$ & 1 & 1 & -1 & -1 & -1 & -1 & 1 & 1\\
$\varepsilon^{\prime}$ & 1 & -1 & 1 & 1 & 1 & -1 & 1 & 1\\
$\varepsilon^{\prime\prime}$ & 1 &  & -1 &  & 1 &  & -1 & \\\hline
\end{tabular}
\end{center}

In the classical case of spin manifolds there is  a relation between the
metric (or spectral) dimension given by the rate of growth of the spectrum of
$D$ and the integer modulo $8$ which appears in the above table. For more
general spaces, however, the two notions of dimension (the dimension modulo $8$
is called the ``$KO$-dimension" because of its origin in $K$-theory) become
independent, since there are spaces $F$ of  metric dimension $0$ but of
arbitrary $KO$-dimension.

The search
to identify the structure of the noncommutative space followed the bottom-up
approach where the known spectrum of the fermionic particles was used to
determine the geometric data that defines the space. 

This bottom-up approach
involved an interesting interplay with experiments. While at first the
experimental evidence of neutrino oscillations contradicted the first attempt, it was realized several years later\footnote{This crucial step was taken independently by John Barrett} in 2006 (see \cite{mc2}),  that the
obstruction to getting neutrino oscillations was naturally eliminated by dropping
the equality between the metric dimension of space-time (which is equal to $4$
as far as we know) and its $KO$-dimension which is only defined modulo $8$.
When the latter is set equal to $2$ modulo $8$  (using the freedom to adjust the geometry of the finite space
encoding the fine structure of space-time) everything works fine: the neutrino
oscillations are there as well as the see-saw mechanism which appears for free
as an unexpected bonus. Incidentally, this also solved the fermion doubling
problem by allowing a simultaneous Weyl-Majorana condition on the fermions to
halve the degrees of freedom.

\subsection{The inner fluctuations of the metric}

In our joint work with A. Chamseddine and W. van Suijlekom \cite{acinner}, we obtained a conceptual understanding of the role of the gauge bosons in physics as the inner fluctuations of the metric. I will describe this result here in a non-technical manner. 

In order to comply with Riemann's requirement that the inverse line element $D$ embodies the forces of nature, it is evidently important that we do not separate artificially the gravitational part from the gauge part, and that $D$ encapsulates both forces in a unified manner. In the traditional geometrization of physics, the gravitational part specifies the metric while the gauge part corresponds to a connection on a principal bundle. In the NCG framework, $D$ encapsulates both forces in a unified manner and the gauge bosons appear as inner fluctuations of the metric but form an inseparable part of the latter. Ignoring at first the important nuance coming from the real structure $J$, the inner fluctuations of the metric were first defined as the transformation 
$$
D\mapsto D+A, \ \ A=\sum a_j[D,b_j], \ \ a_j,  b_j \in \cA, \ A=A^*
$$ 
which imitates the way classical gauge bosons appear as matrix-valued one-forms in the usual framework. The really important facts were that the spectral action applied to $D+A$ delivers the Einstein-Yang-Mills action which combines gravity with matter in a natural manner, and that the gauge invariance becomes transparent at this level since an inner fluctuation coming from a gauge potential of the form $A=u[D,u^*]$ where $u$ is a unitary element (\ie $uu^*=u^*u=1$) simply results in a unitary conjugation $D\mapsto uDu^*$ which does not change the spectral action. 

An equally important fact which emerged very early on, is that as soon as one considers the product of an ordinary geometric space by a finite space of the simplest  nature, such as two points, the inner fluctuations generate the Higgs field and the spectral action gives the desired quartic potential underlying the Brout-Englert-Higgs mechanism. The inverse line element $D_F$ for the finite space $F$ is given by the Yukawa coupling matrix which thus acquires geometric meaning as encoding the geometry of $F$. 

What we discovered in our joint work with A. Chamseddine and W. van Suijlekom \cite{acinner} is that the inner fluctuations arise in fact from the action on metrics (\ie the $D$) of a canonical {\em semigroup} $\pert(\cA)$ which only depends upon the algebra $\cA$ and extends the unitary group.  The semigroup is defined as the self-conjugate elements:
$$
\pert(\cA):=\{A=\sum a_j\otimes b_j^{\rm op}\in  {\mathcal{A}}\otimes {\mathcal{A}}^{\rm op}\mid \sum a_jb_j=1, \ \theta(A)=A\}
$$ 
where $\theta$ is the antilinear automorphism of the algebra ${\mathcal{A}}\otimes {\mathcal{A}}^{\rm op}$ given by 
$$
\theta:\sum a_j\otimes b_j^{\rm op}\mapsto \sum b_j^*\otimes a_j^{*\rm op}.
$$
The composition law in $\pert(\cA)$ is the product in the algebra ${\mathcal{A}}\otimes {\mathcal{A}}^{\rm op}$. The action of this semigroup $\pert(\cA)$ on the metrics is given, for $A=\sum a_j\otimes b_j^{\rm op}$  by 
$$
D\mapsto D'=^{A}\!\!D= \sum a_j D b_j.
$$
Moreover, the transitivity of inner fluctuations results from 
$$
^{A'}\!(^{A}D)=^{(A'A)}\!D.
$$
 What is remarkable is that it allows one to obtain the inner fluctuations in the real case (see  \S\ref{sectreal}), \ie in the presence of the anti-unitary involution $J$, without having to make the ``order one" hypothesis. To do this one uses instead of the algebra $\cA$ the finer one given by $\cB={\mathcal{A}}\otimes \hat {\mathcal{A}} $ where the conjugate algebra $\hat {\mathcal{A}} $ acts in Hilbert space using $JaJ^{-1}$ for $a\in \cA$. The commutation of the actions of ${\mathcal{A}}$ and of $\hat {\mathcal{A}} $  in Hilbert space ensure that $\cB$ acts. One then  simply defines a semigroup homomorphism $\mu: \pert(\cA)\to \pert(\cB)$ by 
 $$
    A\in {\mathcal{A}}\otimes {\mathcal{A}}^{\rm op}\mapsto \mu(A)= A\otimes \hat A\in \left({\mathcal{A}}\otimes \hat {\mathcal{A}}\right)\otimes \left({\mathcal{A}}\otimes \hat {\mathcal{A}}\right)^{\rm op}.
$$
This gives the inner fluctuations in the real case and they take the form 
 \begin{equation*}
D\mapsto D':=D+A_{\left(  1\right)  }+\widetilde{A}_{\left(  1\right)  }+A_{\left(
2\right)  }%
\end{equation*}
where, with $A=\sum a_j\otimes b_j^{\rm op}$ as above 
\begin{align*}
A_{\left(  1\right)  }  &  =%
{\displaystyle\sum\limits_{i}}
a_{i}\left[  D,b_{i}\right] \\
\widetilde{A}_{\left(  1\right)  }  &  =%
{\displaystyle\sum\limits_{i}}
\hat{a}_{i}\left[  D,\hat{b}_{i}\right]  ,\qquad\hat{a}_{i}=Ja_{i}J^{-1}%
,\quad\hat{b}_{i}=Jb_{i}J^{-1}\\
A_{\left(  2\right)  }  &  =%
{\displaystyle\sum\limits_{i,j}}
\hat{a}_{i}a_{j}\left[  \left[  D,b_{j}\right]  ,\hat{b}_{i}\right]  =%
{\displaystyle\sum\limits_{i,j}}
\hat{a}_{i}\left[  A_{\left(  1\right)  },\hat{b}_{i}\right].
\end{align*}
The new quadratic term $A_{\left(  2\right)  }$ vanishes when the order $1$ condition is fulfilled but not in general. This conceptual understanding of the inner fluctuations allowed us, with A. Chamseddine and W. van Suijlekom \cite{acinner, acpati1,acpati2} to determine the inner fluctuations for the natural extension of the Standard Model obtained from the classification of irreducible finite geometries of $KO$-dimension $6$ of \cite{cc5,cc6}. This gives a Pati-Salam extension of the Standard model and we showed in \cite{acpati1,acpati2} that it yields  a natural unification of couplings.

\section{Quanta of Geometry}\label{higherheis}
The above extension of the Standard Model obtained from the classification of irreducible finite geometries of $KO$-dimension $6$ is based on the  finite dimensional algebra $M_2(\H)\oplus M_4(\C)$. While this algebra occurred as one of the simplest in the classification of \cite{cc5,cc6}, its choice remained motivated by the bottom-up approach that we had followed all along up to that point. For instance there was no conceptual explanation for the difference of the real dimensions: 16 for $M_2(\H)$ and 32 for $M_4(\C)$. 

This state of the theory changed drastically in our joint work with A. Chamseddine and S. Mukhanov  \cite{acmu1,acmu2} where the above  finite dimensional algebra $M_2(\H)\oplus M_4(\C)$ appeared unexpectedly from a completely different motivation. The framework is the same, ``spectral geometries" and the question is how to encode all spin Riemannian  $4$-manifolds  in an operator theoretic manner. The key new idea is that since spectral triples only quantize the fundamental $KO$-homology class one should look at the same time for the quantization of the dual $KO$-theory class.

 A hint of this idea can be understood easily in the one dimensional case, \ie for the geometry of the circle. It is an exercise to prove that for unitary representations of the relations
\begin{equation}\label{heisenberg1}
	UU^*=U^*U=1, \ D=D^*, \ \ U^*[D,U]=1
\end{equation}
with $D$ unbounded self-adjoint playing as above the role of the inverse line element, one has\footnote{we refer to \cite{Co-book} for the meaning of the integral symbol} 
\begin{enumerate}
\item $ds$ infinitesimal $\Rightarrow$ $\displaystyle{\int\!\!\!\!\!\! -}\vert ds\vert\in \N$.
\item The formula $d(a,b)=\,{\rm Sup}\,\{\vert f(a)-f(b)\vert\mid \Vert [D,f]\Vert\leq 1\}$ gives the standard distance on the spectrum of $U$ which is the unit circle in $\C$.
\item Let $M$ be a dimension $1$ compact Riemannian manifold, $(\cA,\cH,D)$ the associated spectral triple. Then a solution $U\in \cA$ of the equation $U^*[D,U]=1$ exists if and only if the length $\vert M\vert \in 2\pi \N$.	
\end{enumerate}
One may understand the relations \eqref{heisenberg1} as representations of a group which is a close relative of the Heisenberg group and this would lead one to group representations: but this theme would stay far away from our goal which is 4-dimensional geometries -- and which was achieved in \cite{acmu1,acmu2}.
What we have discovered is a higher geometric analogue
of the Heisenberg commutation relations $[p,q]=i\hbar$. The role of the momentum $p$ is played by the Dirac operator, as amply discussed above. 
The role of the position variable $q$ in the higher analogue of $[p,q]=i\hbar$ was the most difficult to uncover, and another hint was given in \S \ref{ncbonus} where the 2-sphere appeared from very simple non-commuting  discrete variables. The general idea of \cite{acmu1,acmu2}
  is  to encode the analogue of the  position variable $q$ in the same way as the Dirac operator encodes the components of the momenta, just using the Feynman slash. As explained below there are two levels. In the first, which is discussed in \S \ref{sectK}, the quantization is done  for the $K$-theory class, and this justifies the terminology  of $K$-theory higher Heisenberg equation. However, geometrically, the only solutions are disjoint unions of spheres of unit volume. To reach arbitrary compact oriented spin $4$-manifolds, one needs the $KO$-theory refinement. This is treated in \S \ref{sectKO}.

\subsection{The $K$-theory higher Heisenberg equation; Spheres}\label{sectK}

Let us first rewrite the description of the algebra  of \S \ref{ncbonus}, which was presented as $$M_2(\C)\star Y, \  Y=Y^*, \ Y^2=1,\  <Y>=0.$$
As explained in  \S \ref{ncbonus} one can represent its elements as matrices with entries in the commutant of $M_2(\C)$ 
$$
Y=\left(
\begin{array}{cc}
 y_{11} & y_{12} \\
 y_{21} & y_{22} \\
\end{array}
\right)
,\ \ 
Y=\left(
\begin{array}{cc}
 t & z \\
 z^* & -t \\
\end{array}
\right)
$$
where the second form is deduced from the relations. We can rewrite the result in terms of 
$3$   gamma matrices $\Gamma_{A}$, $0\leq A\leq
2$,
$$
\Gamma_{0}=\left(
\begin{array}{cc}
 1 & 0 \\
 0 & -1 \\
\end{array}
\right), \ \ \Gamma_{1}=\left(
\begin{array}{cc}
 0 & 1 \\
 1 & 0 \\
\end{array}
\right), \ \ \Gamma_{2}=\left(
\begin{array}{cc}
 0 & i \\
 -i & 0 \\
\end{array}
\right).
$$
which fulfill:
\[
\left\{  \Gamma_{A},\Gamma_{B}\right\}
=2\,\delta_{AB},\ (\Gamma_{A})^{\ast}=\Gamma_{A}%
\]
and $Y$ now takes the simple form:
$$
Y=Y^{A}\Gamma_{A},\ \ Y^{2}=1,\ \ Y^{\ast}= Y.
$$
\subsubsection{One-sided  higher Heisenberg equation}
This suggests the following extension for arbitrary even $n$.
 We let $Y\in\mathcal{A}\otimes
C_{+}$ be of the Feynman slashed form:
\begin{equation}
Y=Y^{A}\Gamma_{A}, \ Y^{A}\in \cA, \ Y^{2}=1,  \  Y^{\ast}= Y. \label{zzz}%
\end{equation}
Here  $C_{+}\subset M_{s}(\mathbb{C})$, $s=2^{n/2}$, is an irreducible representation of 
the Clifford algebra on $n+1$ gamma matrices $\Gamma_{A}$, $0\leq A\leq
n$
\[
\Gamma_{A}\in C_{+}, \quad\left\{  \Gamma_{A},\Gamma_{B}\right\}
=2\,\delta_{AB},\ (\Gamma_{A})^{\ast}=\Gamma_{A}.%
\]
The one-sided higher analogue of the Heisenberg commutation relations is
\begin{equation}
\frac{1}{n!}\left\langle Y\left[  D,Y\right]^n
\right\rangle =\,\gamma  \label{yyy}%
\end{equation}
where the notation $\left\langle T\right\rangle $ means the \emph{normalized}
trace of $T=T_{ij}$ with respect to the above matrix algebra $M_{s}%
(\mathbb{C})$ ($1/s$ times the sum of the $s$ diagonal terms $T_{ii}$). 
\subsubsection{Quantization of volume}\label{quant volume}
For even $n,$ equation \eqref{yyy}, together with
the hypothesis that the eigenvalues of $D$ grow as in dimension $n$ (\ie that $ds$ is an infinitesimal of order $1/n$) imply
that the volume, expressed as the leading term in the Weyl asymptotic formula
for counting eigenvalues of the operator $D$, is {\em 
quantized} by being equal to the index pairing of the
operator $D$ with the $K$-theory class of $\mathcal{A}$ defined by (note that $s$ is even)
 $$[e-1/2]:= [e]-s/2 [1_{\cA}]\in K_0(\cA), \ \ e=(1+\,Y)/2.$$
 To understand this result, we need to recall that the integral pairing between $K$-homology and $K$-theory is computed by the pairing of the Chern characters in cyclic theory according to the diagram:
 {\bf\color{blue}
$$
\xymatrix@C=45pt@R=45pt{
 K-\text{Theory} \ \ar[d]_{Ch_*}\ar@{<->}[rr] && K-\text{Homology} \ar@{->>}[d]^{Ch^*}
\\
HC_*\ \ar@{<->}[rr]&& HC^*}
$$}
 While the Chern character from $K$-homology to cyclic cohomology is difficult, its counterpart from $K$-theory to cyclic homology can be explained succinctly as follows. Given a unital (not assumed commutative) algebra $\cA$, the $(b,B)$-bicomplex is obtained from the $(b,B)$ bicomplex:
$$\underline{\cA}:=\cA/\C1, \ \underline{C}_n(\cA):=\cA\otimes \underline{\cA}\otimes \cdots \otimes \underline{\cA}
$$
$$
b(a_0\otimes \cdots \otimes a_n):=a_0a_1\otimes \cdots \otimes a_n-a_0\otimes a_1a_2\otimes \cdots \otimes a_n +\cdots +
$$
$$
(-1)^{n-1}a_0\otimes \cdots \otimes a_{n-1} a_n+(-1)^{n}a_n a_0\otimes \cdots \otimes a_{n-1}
$$
$$
B(a_0\otimes \cdots \otimes a_n):=\sum_0^n (-1)^{nj} 1\otimes a_j\otimes a_{j+1}\otimes \cdots \otimes a_{j-1}.
$$
The operations $(b,B)$ fulfill $$
b^2=0,\ B^2=0, \ bB=-Bb
$$ 
and an even (resp. odd) cycle $c=(c_n)$ is given by its components $c_n\in \underline{C}_n(\cA)$ for $n$ even (resp. odd) which fulfill 
\begin{equation}\label{cycle}
	Bc_n+bc_{n+2}=0 \qqq n \ \text{even} \ \ \text{(resp.}\ \text{odd)}.
\end{equation}

The Chern character of an idempotent $e\in \cA$, $e^2=e$,  is then given by the cycle with components 
$\mathrm{Ch}_{0}(e)=e\in \cA=\underline{C}_0(\cA)$ and for $k>0$
$$
\mathrm{Ch}_{2k}(e)=\lambda_k \times (e-\frac 12)\otimes e \otimes e \otimes \cdots  \otimes e  \in \underline{C}_{2k}(\cA).
$$
One has 
$$
b(e-\frac 12)\otimes e \otimes e \otimes \cdots  \otimes e=\frac 12 \left(1\otimes e \otimes e \otimes \cdots  \otimes e \right)
$$
$$
B(e-\frac 12)\otimes e \otimes e \otimes \cdots  \otimes e=B(e\otimes e \otimes e \otimes \cdots  \otimes e)
$$
$$
=(2k+1)\left(1\otimes e \otimes e \otimes \cdots  \otimes e \right).
$$
Thus one can choose the $\lambda_k$ so that $(2k+1)\lambda_k+\frac 12 \lambda_{k+1}=0$ 
$$
B\mathrm{Ch}_{2k}(e)+b\mathrm{Ch}_{2k+2}(e)=0
$$
and one gets a cycle $\mathrm{Ch}_{*}(e)$ in the $(b,B)$-bicomplex which gives the Chen character in $K$-theory.

In general the idempotent $e$ does not belong to $\cA$ but to matrices $M_{s}(\mathcal{A})$ and 
the next step is to pass to matrices. To do this one considers partial trace maps 
$$
\mathrm{tr}:\underline{C}_n(M_{s}(\mathcal{A}))\to \underline{C}_n(\cA).
$$
One defines 
\[
\mathrm{tr}:M_{s}(\mathcal{A})\otimes M_{s}(\mathcal{A})\otimes\cdots\otimes
M_{s}(\mathcal{A})\rightarrow\mathcal{A}\otimes\mathcal{A}\otimes\cdots
\otimes\mathcal{A}%
\]
as the linear map such that:
$$
\mathrm{tr}\left(  (a_{0} \otimes\mu_{0})\otimes(a_{1}\otimes\mu_{1}%
)\otimes\cdots\otimes(a_{m}\otimes\mu_{m})\right)  = \mathrm{Trace}(\mu
_{0}\cdots\mu_{m})\ a_{0}\otimes a_{1}\otimes\cdots\otimes a_{m}%
$$
where $\mathrm{Trace}$ is the ordinary trace of matrices. 
Let us denote by
$\iota_{k}$ the operation which inserts a $1$ in a tensor at the $k$-th place.
So for instance
\[
\iota_{0}(a_{0}\otimes a_{1}\otimes\cdots\otimes a_{m})=1\otimes a_{0}\otimes
a_{1}\otimes\cdots\otimes a_{m}%
\]
One has $\mathrm{tr}\circ\iota_{k}=\iota_{k}\circ\mathrm{tr}$ since (taking
$k=0$ for instance)
\[
\mathrm{tr}\circ\iota_{0}\left(  (a_{0}\otimes\mu_{0})\otimes(a_{1}\otimes
\mu_{1})\otimes\cdots\otimes(a_{m}\otimes\mu_{m})\right)  =
\]%
\[
=\mathrm{tr}\left(  (1\otimes1)\otimes(a_{0}\otimes\mu_{0})\otimes
(a_{1}\otimes\mu_{1})\otimes\cdots\otimes(a_{m}\otimes\mu_{m})\right)
\]%
\[
=\mathrm{Trace}(1\mu_{0}\cdots\mu_{m})1\otimes a_{0}\otimes a_{1}\otimes
\cdots\otimes a_{m}=
\]%
\[
=\iota_{0}\left(  \mathrm{tr}\left(  (a_{0}\otimes\mu_{0})\otimes(a_{1}%
\otimes\mu_{1})\otimes\cdots\otimes(a_{m}\otimes\mu_{m})\right)  \right).
\]
Thus the map $\tr$ induces a map $\mathrm{tr}:\underline{C}_n(M_{s}(\mathcal{A}))\to \underline{C}_n(\cA)$ and one checks that this map is compatible with the operations $(b,B)$. For an idempotent $e\in M_{s}(\mathcal{A})$ the components of its Chern character in $\underline{C}_*(\cA)$ are given by $\tr(\mathrm{Ch}_{2k}(e))$. Thus  they are 
$$
\mathrm{Ch}_{2k}(e)=\lambda_k \times \mathrm{tr}\left((e-\frac 12)\otimes e \otimes e \otimes \cdots  \otimes e\right), \ \ k>0. 
$$
Moreover this formula still holds  for $k=0$ when replacing $e$ by $[e-1/2]$:
$$
\mathrm{Ch}_{2k}([e-1/2])=\lambda_k \times \mathrm{tr}\left((e-\frac 12)\otimes e \otimes e \otimes \cdots  \otimes e\right) \qqq k\geq 0.
$$
Using $Y=2e-1$ and the construction of the $\underline{C}_m(\cA)$ one thus gets, for $m=2k$ even, 
\begin{equation}\label{phim}
	\mathrm{Ch}_{m}([e-1/2])=2^{-(m+1)}\lambda_k\mathrm{tr}\left(  Y\otimes Y\otimes Y\otimes
\cdots\otimes Y\right)  \in\underline{C}_m(\cA). 
\end{equation}
The fundamental fact which is behind the quantization of the volume is, for $Y$ fulfilling \eqref{zzz}, the vanishing of all the lower components 
\begin{equation}\label{vanishing lower}
	\mathrm{Ch}_{m}([e-1/2])=0, \ \forall m<n.
\end{equation}
This follows because for a product $P$ of an odd number $2k+1<n+1$ of $\Gamma_A$, the trace of $P$ vanishes since one can still find a $\Gamma=\Gamma_X$ which anti-commutes with the $\Gamma_A$'s involved in $P$ and thus 
$$
P=\Gamma^2 P=-\Gamma P \Gamma\Rightarrow {\rm Trace}(P)=0. 
$$
It follows from \eqref{vanishing lower} that the component $\mathrm{Ch}_{n}([e-1/2])$ is a Hochschild cycle and
that for any cyclic $n$-cocycle $\phi_{n}$ the pairing $<\phi_{n},e>$ is the
same as $<I(\phi_{n}),\mathrm{Ch}_{n}(e)>$ where $I(\phi_{n})$ is the
Hochschild class of $\phi_{n}$. This applies to the cyclic $n$-cocycle
$\phi_{n}$ which is the Chern character $\phi_{n}$ in $K$-homology of the
spectral triple $(\mathcal{A},\mathcal{H},D)$ with grading $\gamma$ where
$\mathcal{A}$ is the algebra generated by the components $Y^{A}$ of $Y$. One then uses the following formula for the Hochschild class $\tau$ of the Chern character $\phi_{n}$ in $K$-homology of the
spectral triple $(\mathcal{A},\mathcal{H},D)$, up to normalization:\footnote{we refer to \cite{Co-book} for the meaning of the integral symbol}
\[
\tau(a_{0},a_{1},\ldots,a_{n})={\int\!\!\!\!\!\!-}\gamma a_{0}[D,a_{1}%
]\cdots\lbrack D,a_{n}]D^{-n},\ \ \forall a_{j}\in\mathcal{A}.
\]
This follows from the local index formula of Connes-Moscovici \cite{cmindex}. But in fact, one does
not need the technical hypothesis  since, when the lower
components of the operator theoretic Chern character all vanish, one can use
the non-local index formula in cyclic cohomology and the determination in
the  book \cite{Co-book} of the Hochschild class of the
index cyclic cocycle. We refer to \cite{carey} for an optimal formulation of the result. 
Moreover since $D$  commutes with the algebra $M_s(\C)$ one has
$$
\tau\circ \tr (y_{0},y_{1},\ldots,y_{n})=s{\int\!\!\!\!\!\!-}\gamma \left\langle y_{0}[D,y_{1}%
]\cdots\lbrack D,y_{n}]\right\rangle D^{-n},\ \ \forall y_{j}\in M_s(\mathcal{A})
$$
 so that with $Y=2e-1$, one gets
\[
<\tau,\mathrm{Ch}_{n}([e-1/2])>=s{\int\!\!\!\!\!\!-}\gamma\left\langle Y\left[
D,Y\right]  ^{n}\right\rangle D^{-n}%
\]
and, up to normalization, equation \eqref{yyy} thus implies
$$
{\int\!\!\!\!\!\!-}D^{-n}={\int\!\!\!\!\!\!-}\gamma^2 D^{-n}=\frac{1}{n!}
{\int\!\!\!\!\!\!-}\gamma\left\langle Y\left[
D,Y\right]  ^{n}\right\rangle D^{-n}=\frac{1}{n!s}<\tau,\mathrm{Ch}_{n}([e-1/2])>\in \frac{1}{n!s}\ \Z:
$$
which is the quantization of the volume.
\subsubsection{Disjoint Quanta}
We recall that given a smooth compact oriented spin manifold $M$, the
associated spectral triple $(\mathcal{A},\mathcal{H},D)$ is given by the
action in the Hilbert space $\mathcal{H}=L^{2}(M,S)$ of $L^{2}$-spinors of the
algebra $\mathcal{A}=C^{\infty}(M)$ of smooth functions on $M$, and the Dirac
operator $D$ which in local coordinates is of the form
$$
D=\gamma^{\mu}\left(  \frac{\partial}{\partial x^{\mu}}+\omega_{\mu}\right)
$$
where $\gamma^{\mu}=e_{a}^{\mu}\gamma^{a}$ and $\omega_{\mu}$ is the spin-connection
\begin{thm}
Let $M$ be a spin Riemannian manifold of even dimension $n$ and
$(\mathcal{A},\mathcal{H},D)$ the associated spectral triple. Then a solution
of the one-sided equation \eqref{yyy} exists if and only if $M$ decomposes as the disjoint sum
of spheres of unit volume. On each of these irreducible components the unit
volume condition is the only constraint on the Riemannian metric which is
otherwise arbitrary for each component.	
\end{thm}
\begin{figure}[H]
\begin{center}
\includegraphics[scale=0.6]{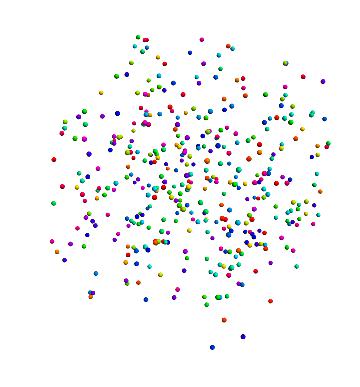}
\end{center}
\caption{Collection of tiny spheres \label{tiny} }
\end{figure}
 Equation \eqref{zzz} shows that a solution
$Y$ gives a map $Y:M\rightarrow S^{n}$ from the
manifold $M$ to the $n$-sphere. Let us compute the left hand side of \eqref{yyy}. The normalized trace of the
product of $n+1$ Gamma matrices is the totally antisymmetric tensor
\[
\left\langle \Gamma_{A}\Gamma_{B}\cdots\Gamma_{L}\right\rangle =i^{n/2}%
\epsilon_{AB\ldots L},\ \ A,B,\ldots,L\in\{1,\ldots,n+1\}.
\]
One has $$\left[  D,Y\right]  =\gamma^{\mu}\frac{\partial Y^{A}}{\partial
x^{\mu}}\Gamma_{A}=\nabla Y^{A}\Gamma_{A}$$ where we let $\nabla f$ be the
Clifford multiplication by the gradient of $f$. Thus one gets at any $x\in M$
the equality
$$
\left\langle Y\left[  D,Y\right]  \cdots\left[  D,Y\right]  \right\rangle
=i^{n/2}\epsilon_{AB\ldots L}Y^{A}\nabla Y^{B}\cdots\nabla Y^{L}.%
$$
Given $n$ operators $T_{j}\in\mathcal{C}$ in
an algebra $\mathcal{C}$ the multiple commutator
\[
\lbrack T_{1},\ldots,T_{n}]:=\sum\epsilon(\sigma)T_{\sigma(1)}\cdots
T_{\sigma(n)}
\]
(where $\sigma$ runs through all permutations of $\{1,\ldots,n\}$) is a
multilinear totally antisymmetric function of the $T_{j}\in\mathcal{C}$. In
particular, if the $T_{i}=a_{i}^{j}S_{j}$ are linear combinations of $n$
elements $S_{j}\in\mathcal{C}$ one gets
\begin{equation}\label{antisym}
\lbrack T_{1},\ldots,T_{n}]=\mathrm{Det}(a_{i}^{j})[S_{1},\ldots
,S_{n}].
\end{equation}
For fixed $A$, and $x\in M$ the sum over the other indices
\[
\epsilon_{AB\ldots L}Y^{A}\nabla Y^{B}\cdots\nabla Y^{L}=(-1)^{A}Y^{A}[\nabla
Y^{1},\nabla Y^{2},\ldots,\nabla Y^{n+1}]
\]
where all other indices are $\neq A$. At $x\in M$ one has $\nabla Y^{j}
=\gamma^{\mu}\partial_{\mu}Y^{j}$ and by \eqref{antisym} the multi-commutator
(with $\nabla Y^{A}$ missing) gives
\[
\lbrack\nabla Y^{1},\nabla Y^{2},\ldots,\nabla Y^{n+1}]=\epsilon^{\mu\nu
\ldots\lambda}\partial_{\mu}Y^{1}\cdots\partial_{\lambda}Y^{n+1}[\gamma
^{1},\ldots,\gamma^{n}].
\]
Since $\gamma^{\mu}=e_{a}^{\mu}\gamma_{a}$ and $i^{n/2}[\gamma_{1}
,\ldots,\gamma_{n}]=n!\gamma$ one thus gets 
$$
\left\langle Y\left[  D,Y\right]  \cdots\left[  D,Y\right]  \right\rangle
=n!\gamma\mathrm{Det}(e_{a}^{\alpha})\,\omega\label{expectY}%
$$
where
\[
\omega=\epsilon_{AB\ldots L}Y^{A}\partial_{1}Y^{B}\cdots\partial_{n}Y^{L}%
\]
so that $\omega dx_{1}\wedge\cdots\wedge dx_{n}$ is the pullback $Y^{\#}%
(\rho)$ by the map $Y:M\rightarrow S^{n}$ of the rotation invariant volume
form $\rho$ on the unit sphere $S^{n}$ given by
\[
\rho=\frac{1}{n!}\epsilon_{AB\ldots L}Y^{A}dY^{B}\wedge\cdots\wedge dY^{L}.%
\]
Thus, using the inverse vierbein, the one-sided equation \eqref{yyy} is
equivalent to
$$
\det\left(  e_{\mu}^{a}\right)  dx_{1}\wedge\cdots\wedge dx_{n}=Y^{\#}
(\rho). 
$$
This equation  implies that the Jacobian of the map
$Y:M\rightarrow S^{n}$ cannot vanish anywhere, and hence that the map $Y$ is a
covering.

It would seem at this point that only disconnected geometries fit in this
framework.  But this would be to  ignore an essential piece of structure of the NCG
framework, which allows one to refine \eqref{yyy}. Namely: the real structure
$J$, an antilinear isometry in the Hilbert space $\cH$ which is the algebraic
counterpart of charge conjugation.

\subsection{The $KO$-theory higher Heisenberg equation}\label{sectKO}
We now take into account the real structure 
 $J$ and this gives the refinement  from  $K$ to $KO$.  One 
 replaces \eqref{zzz} by (with summation on indices $A$ and $\kappa\in \{\pm 1\}$) 
 \begin{equation}\label{xxx}
 	Y=Y^{A}_{\kappa}\Gamma_{A,\kappa}, \ Y^4=1,\ Y^*Y=1, 
 \end{equation}
The Hilbert space splits according to the spectrum of $Y^2$ as a direct sum $\cH=\cH(+)\oplus \cH(-)$ 
$$
Y=Y_{+}\oplus Y_{-}, \ \ Y_{\pm}^2=\pm 1, \ \ Y_{\pm}^*=\pm Y_{\pm},
\ \
Y_{\pm}=Y^{A}_{\pm}\Gamma_{A,\pm}.
$$
For $\kappa\in \{\pm 1\}$ the $\Gamma_{A,\kappa}$ fulfill in $\cH(\kappa)$ the Clifford relations
$$
\left\{  \Gamma_{A,\kappa},\Gamma_{B,\kappa}\right\}
=2\kappa\,\delta_{AB},\ (\Gamma_{A,\kappa})^{\ast}=\kappa\Gamma_{A,\kappa}.
$$
The compatibility with  $J$ is given by the relations:
 $$
 JY^2=-Y^2J, \ [Y,JYJ^{-1}]=0
 $$
Let $C_{\pm}=C(\Gamma_{A,\pm})$ be the algebra generated over $\R$ by the $\Gamma_{A,\kappa}$. 
In $\cH(+)$, $C_+$ commutes with $C'_-=JC_-J^{-1}$ to take into account the relation $[Y,JYJ^{-1}]=0$. We thus view $Y$ as a smooth section:
 $$Y=Y_{+}\oplus Y_{-}\in C^{\infty}(M,C_{+}\oplus C_{-})$$
This leads us to refine the quantization
condition by taking $J$ into account as the two-sided equation

\begin{equation}
\frac{1}{n!}\left\langle Z\left[  D,Z\right]^n  \right\rangle
=\gamma,\quad Z=2EJEJ^{-1}-1,\ [D,Y^2]=0,\label{jqq}
\end{equation}
where $E$ is the spectral projection $E_{+}\oplus E_{-}$ of the unitary 
$Y=Y_{+}\oplus Y_{-}\in C^{\infty}(M,C_{+}\oplus C_{-})$.   $$E=E_{+}\oplus E_{-}=\frac{1}{2}(1+Y_{+})\oplus\frac{1}{2}(1+iY_{-})$$

It turns out  that in dimension $n=4$,  the irreducible pieces  give :
$$C_{+}=M_{2}(\mathbb{H}),\  \  \  C_{-}=M_{4}(\mathbb{C})$$
 which give the algebraic constituents of the Standard Model
exactly in the form of our previous work. This can be seen using the following table:

\bigskip 

\begin{center}
\begin{tabular}{|c|c||c|c|}
  \hline
  $p-q$  &  Cliff$_{p,q}(\R)$&$p-q$  &  Cliff$_{p,q}(\R)$\\
  mod $8$ & $n=p+q$& mod $8$& $n=p+q$\\
   \hline
    & &&\\
 0 &  $M(2^{n/2},\R)$& 1 &$M_u(\R)\oplus M_u(\R)$ \\
 &&& $u=2^{(n-1)/2}$\\
  \hline
  & &&\\
  2  &  $M(2^{n/2},\R)$& 3&$M(2^{(n-1)/2},\C)$ \\
  &&&\\
  \hline
  &&& \\
 4&     $M(2^{(n-2)/2},\H)$& 5& $M_v(\H)\oplus M_v(\H)$\\
    &&& $v=2^{(n-3)/2}$\\
  \hline
  & & &\\
  6 &   $M(2^{(n-2)/2},\H)$ & 7 & $M(2^{(n-1)/2},\C)$\\
  &&&\\
    \hline
    \end{tabular}
\bigskip
\end{center}
Indeed in dimension $n=4$  one needs $n+1=5$ gamma matrices, and the irreducible pieces of the Clifford algebras Cliff$_{p,q}(\R)$ are $M_{2}(\mathbb{H})$ for $(p,q)=(5,0)$ and $M_{4}(\mathbb{C})$ for $(p,q)=(0,5)$. Moreover 
in the $4$-dimensional case one has, in the Hilbert space $\cH(+)$, by the detailed calculation of \cite{acmu2},
\[
\left\langle Z\left[  D,Z\right]  ^{4} \right\rangle_{+} =\frac{1}{2}\left\langle
Y_{+}\left[  D,Y_{+}\right]  ^{4} \right\rangle +\frac{1}{2}\left\langle Y_{-}^{\prime
}\left[  D,Y_{-}^{\prime}\right]  ^{4} \right\rangle ,
\]
where $Y'_{-}=i J Y_{-}J^{-1}$. 
One
now gets two maps $Y_{\pm}:M\rightarrow S^{n}$ while  \eqref{jqq}
becomes, up to normalization
\begin{equation}
\det\left(  e_{\mu}^{a}\right)  dx_{1}\wedge\cdots\wedge dx_{n}  =Y^{\#}_{+}(\rho)+Y^{\#}_{-}(\rho), \label{qqq}
\end{equation}
where $Y^{\#}_{\pm}(\rho)$ is the pull back of the  volume   form $\rho$
of the sphere.

For an $n$-dimensional smooth compact manifold we let 
$D(M)$ be the set of pairs of smooth maps
$\phi_{\pm}:M\rightarrow S^{n}$ such that the differential form
\[
\phi_{+}^{\#}(\rho)+\phi_{-}^{\#}(\rho)=\omega
\]
does not vanish anywhere on $M$ 
($\rho$ is the standard volume form on the sphere $S^{n}$).
\begin{defn} Let $M$ be an $n$-dimensional oriented smooth compact manifold
\[
q_M:=\{\mathrm{deg}(\phi_{+})+\mathrm{deg}(\phi_{-})\mid(\phi_{+}%
,\phi_{-})\in D(M)\}
\]
where $\mathrm{deg}(\phi)$ is the topological degree of 
$\phi$.	
\end{defn}

\begin{thm}\label{emerge}
$(i)$~Let $M$ be a compact oriented spin Riemannian manifold of dimension
$4$. Then a solution of \eqref{qqq} exists if and only if the volume of $M$ is
quantized to belong to the invariant $q_{M}\subset\mathbb{Z}$.\newline
$(ii)$~Let $M$ be a smooth connected oriented compact spin $4$-manifold.
Then  $q_M$ contains all integers $m\geq5$.		
\end{thm}
 The invariant $q_{M}$ makes sense in any dimension. For $n=2,3$, and any $M$,
it contains all sufficiently large integers. The case $n=4$ is more difficult;
but we showed in \cite{acmu2} that for any Spin manifold it contains all integers $m>4$.
This uses fine results on the existence of ramified covers of the sphere and on
immersion theory going back to Smale, Milnor and Poenaru.
By a result of M. Iori, R. Piergallini \cite{piergallini},
any orientable closed (connected) smooth 4-manifold is a simple
5-fold cover of $S^4$ branched over a smooth surface (meaning that the covering map can
be assumed to be smooth).
The key lemma\footnote{I am indebted to Simon Donaldson for his generous help in finding this key result} which allows one to then rely on immersion theory and apply the fundamental result of Poenaru \cite{Poenaru} (on the existence of an immersion in $\R^n$ of any open parallelizable $n$-manifold) is the following:
\begin{lem}
Let $\phi:M\to S^{4}$ be a smooth map such that $\phi
^{\#}(\alpha)(x)\geq0$ $\forall x\in M$ and let $R=\{x\in M\mid\phi^{\#}
(\alpha)(x)=0\}$. Then there exists a map $\phi^{\prime}$ such that $\phi
^{\#}(\alpha)+\phi^{\prime\#}(\alpha)$ does not vanish anywhere if and only if
there exists an immersion $f:V\to\mathbb{R}^{4}$ of a neighborhood $V$ of $R$.
Moreover if this condition is fulfilled one can choose $\phi^{\prime}$ to be
of degree $0$.	
\end{lem}

The spin condition on the $4$-manifold allows one to prove that the neighborhood $V$ is parallelizable. By a result of A. Haefliger, the spin condition is equivalent to the vanishing of the second Stiefel-Whitney class $w_2$ of the tangent bundle.
In the converse direction,   
Jean-Claude Sikorav and Bruno Sevennec found the following obstruction which
implies for instance that $D(\mathbb{C}P^{2})=\emptyset$. 
Let $M$ be an oriented compact smooth $4$-dimensional manifold, then, with
$w_{2}$ the second Stiefel-Whitney class of the tangent bundle,
\[
D(M)\neq\emptyset\implies w_{2}^{2}=0
\]
Indeed if $D(M)\neq\emptyset$ one  has a cover of $M$ by two open sets on which the tangent bundle is stably
trivialized. Thus the above product of two Stiefel-Whitney classes vanishes.
\begin{figure}[H]
\begin{center}
\includegraphics[scale=1]{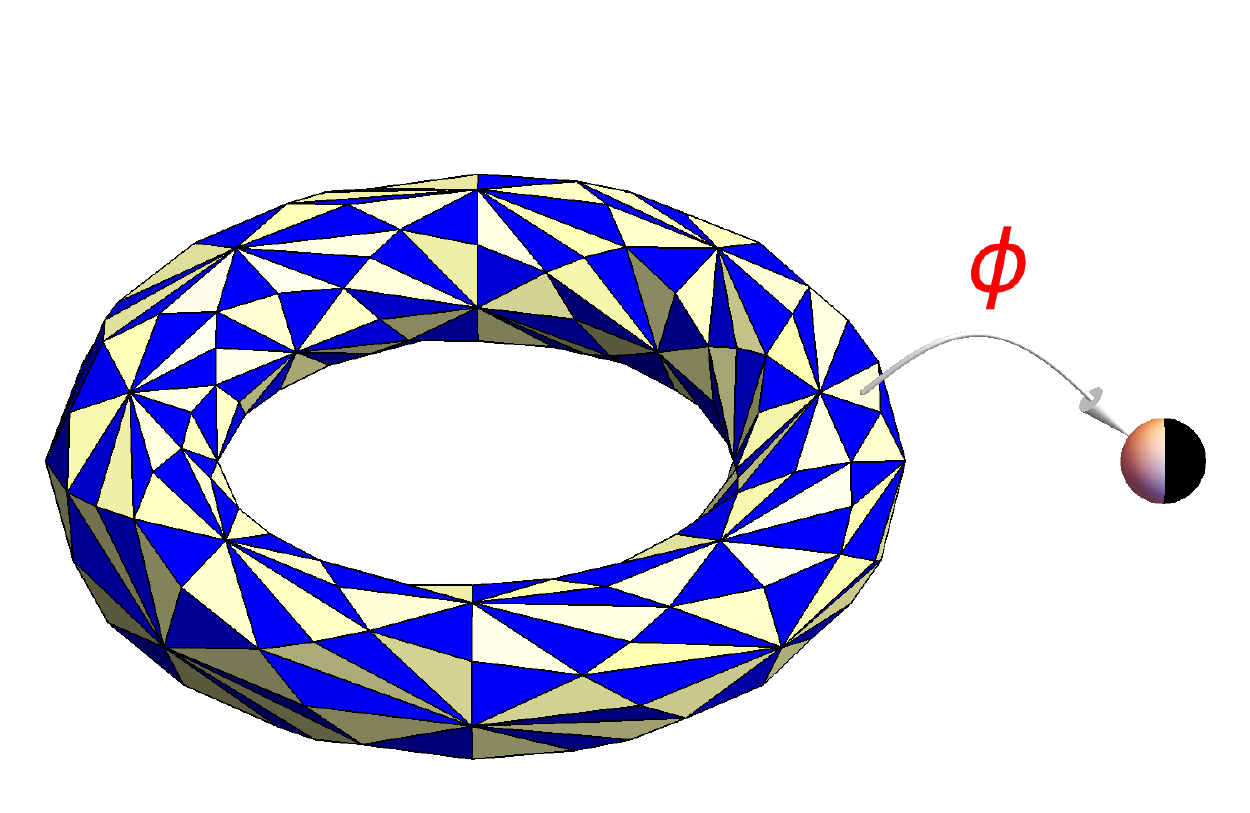}
\end{center}
\caption{Ramified cover of the sphere \label{butterfly} }
\end{figure}
\subsection{Emerging Geometry}\label{emerging geom}
 Theorem \ref{emerge} shows how $4$-dimensional spin geometries arise from irreducible representations of simple algebraic relations. There is no restriction to fix the Hilbert space $\cH$ as well as the actions of the Clifford algebras $C_\pm$ and of $J$ and $\gamma$. The remaining indeterminate
 operators  are $D$ and $Y$. They fulfill equation \eqref{jqq}. The geometry appears from the joint spectrum of the $Y^A_\pm$ and is a $4$-dimensional immersed submanifold in the $8$-dimensional product $S^4\times S^4$. 
Thus this suggests taking the
operators $Y, D$  as being the correct variables for a first shot
at a theory of quantum gravity. 
In the sequel the algebraic relations between $Y_{\pm}$, $D$, $J$,
$C_{\pm}$, $\gamma$ are assumed to hold.
As we have seen above  a compact spin $4$-dimensional manifold $M$ appears as immersed by  a map
$(Y_+,Y_-):M \to S^{4}\times S^{4}$.
 An interesting question which comes in this
respect is whether, given a compact spin $4$-dimensional manifold $M$, one can find a map
$(Y_+,Y_-):M \to S^{4}\times S^{4}$ which embeds $M$ as a submanifold of
$S^{4}\times S^{4}$. One has the 
strong Whitney embedding theorem: $M^4\subset \R^4\times \R^4\subset S^4\times S^4$
so there is no a-priori obstruction to expect an embedding rather than an immersion.
It is worthwhile to mention that a generic immersion would in fact suffice to reconstruct the manifold.  Next, in general, if one starts from a representation of the algebraic relations, there are two  natural questions:
\begin{enumerate}
\item[A):] Is it true that the joint spectrum of the $Y_+^{A}$ and $Y_-^{
B}$ is of dimension $4$ while one has $8$ variables?

\item[B):] Is it true that the volume
$
{\int\!\!\!\!\!\! -} D^{-4}
$
remains quantized?
\end{enumerate}
\subsubsection{Dimension}
The reason why $A)$ holds in the case of classical manifolds is that in that
case the joint spectrum of the $Y^{A}$ and $Y^{\prime B}$ is the subset of
$S^{4}\times S^{4}$ which is the image of the manifold $M$ by the map $x\in
M\mapsto(Y(x),Y^{\prime}(x))$ and thus its dimension is at most $4$.

The reason why $A)$ holds in general is because of the assumed boundedness of
the commutators $[D,Y]$ and $[D,Y^{\prime}]$ together with the commutativity
$[Y,Y^{\prime}]=0$ (order zero condition) and the fact that the spectrum of
$D$ grows like in dimension $4$.

\subsubsection{Quantization of volume}
The reason why $B)$ holds in the general case is that the results of \S \ref{quant volume} apply separately to $Y_{+}$ and $Y_{-}^{\prime}$. This gives, up to a normalization constant $c_4\neq 0$, the integrality
\[
{\int\!\!\!\!\!\! -} \gamma\left\langle Y_{+}\left[  D,Y_{+}\right]
^{4}\right\rangle D^{-4}=c_4 <[D],[e-1/2]>\in c_4\,\Z
\]
\[
{\int\!\!\!\!\!\! -} \gamma\left\langle Y_{-}^{\prime
}\left[  D,Y_{-}^{\prime
}\right]
^{4}\right\rangle D^{-4}=c_4 <[D],[e'-1/2]>\in c_4\, \Z.
\]
Thus the equality
\[
\left\langle Z\left[  D,Z\right]  ^{4} \right\rangle_{+} =\frac{1}{2}\left\langle
Y_{+}\left[  D,Y_{+}\right]  ^{4} \right\rangle +\frac{1}{2}\left\langle Y_{-}^{\prime
}\left[  D,Y_{-}^{\prime}\right]  ^{4} \right\rangle .
\]
together with equation \eqref{jqq} gives in $\cH(+)$,
$$
\frac{1}{2}\left\langle
Y_{+}\left[  D,Y_{+}\right]  ^{4} \right\rangle +\frac{1}{2}\left\langle Y_{-}^{\prime
}\left[  D,Y_{-}^{\prime}\right]  ^{4} \right\rangle=4!\gamma
$$
and one gets from $\gamma^2=1$:
\begin{thm} In any operator representation of the two sided equation \eqref{jqq} in
which the spectrum of $D$ grows as in dimension $4$ the volume (the leading
term of the Weyl asymptotic formula) is quantized, 
 (up to a normalization constant $c>0$)
\[
{\int\!\!\!\!\!\! -} D^{-4}\in c\, \mathbb{N }.
\]	
\end{thm}
This quantization of the volume implies that the bothersome cosmological leading term of the spectral action is now quantized; and thus it no longer appears in the differential variation of the spectral action.  Thus and provided one understands better how to reinstate all the fine details of the finite geometry (the one encoded by the Clifford algebras)  the variation of the spectral action will reproduce the Einstein equations
coupled with matter.
\subsection{Final remarks} Finally, we briefly discuss a few important points which would require more work of clarification if one wants to get a bit closer to the goal of unification at the ``pre-quantum" level,  best described in Einstein's words (see H. Nicolai, Cern Courier, January 2017) as follows:

``Roughly but truthfully, one might say: we not only want to understand how nature works, but we are also after the perhaps utopian and presumptuous goal of understanding why nature is the way it is and not otherwise."
\begin{enumerate}
\item All our discussion of geometry takes place in the Euclidean signature. Physics takes place in the Minkowski signature. The Wick rotation plays a key role in giving a mathematical meaning to the Feynman integral in QFT for flat space-time  but becomes problematic for curved space-time. But following Hawking and Gibbons one can investigate the Euclidean Feynman integral over compact $4$-manifolds implementing a cobordism between two fixed $3$-geometries. Two interesting points occur if one uses the above spectral approach. First the new boundary terms, involving the extrinsic curvature of the boundary, which Hawking and Gibbons had to add to the Einstein action, pop up automatically from the spectral action: as shown in \cite{cc7}. Second, in the functional integral, the kinetic term of the Weyl term (\ie the ``dilaton") has the wrong sign. In our formalism the higher Heisenberg equation fixes the volume form and automatically freezes the dilaton.

\item The number of generations is not predicted by the above theory. The need to have this multiplicity in the representation of the finite algebra $\cA_F$ might  be related to the discussion of \S\ref{notionof manifold} in the following way. For non-simply connected spaces the Poincar\' e duality $KO$-fundamental class should take into account the fundamental group. We skipped over this point in \S\ref{notionof manifold}; and in the non-simply connected case one needs to twist the fundamental $KO$-homology class by flat bundles. It is conceivable that the generations appear from such a twist by a $3$-dimensional representation. This could be a good motivation to extend the classical treatment of flat bundles (\ie of representations of the fundamental group) to the general case of noncommutative spaces.

\end{enumerate}

\section{Appendix}\label{app1}

Here is a possible translation of the second quote of Grothendieck:

{\em

It must be already fifteen or twenty years ago that, leafing through the modest volume constituting the complete works
of Riemann, I was struck by a remark of his ``in passing". He pointed out that it could well be that the ultimate structure of space is discrete, while the continuous representations that we
 make of it constitute perhaps a simplification (perhaps excessive, in the long run ...) of a more complex reality; That for the human mind, ``the continuous"  was easier to grasp than ``the discontinuous",  and that it serves us, therefore, as an ``approximation" to apprehend the discontinuous.

This is a remark of a surprising penetration in the mouth of a mathematician, at a time when the Euclidean model of physical space had never yet been questioned; in the strictly logical sense, it is rather the discontinuous which traditionally served as a mode of technical approach to the continuous.

Mathematical developments of recent decades have, moreover, shown a much more intimate symbiosis between continuous and discontinuous structures than was  imagined, even in the first half of this century.

In any case finding a ``satisfactory" model (or, if necessary, a set of such models, ``satisfactorily connecting" to each other) of ``continuous",``discrete" or of ``mixed" nature - such work will surely involve a great conceptual imagination, and a consummate flair for apprehending and unveiling new type mathematical structures.

This kind of imagination or ``flair" seems rare to me, not only among physicists (where Einstein and Schr\"odinger seem to have been among the rare exceptions), but even among mathematicians (and here I speak with full knowledge).

To summarize I predict that the expected renewal (if it must yet  come) will  come from a mathematician in soul well informed about  the great problems of physics, rather than from a physicist. But above all, it will take a man with ``philosophical openness" to grasp the crux of the problem. This is by no means a technical one but rather a fundamental problem of natural philosophy."}

\end{document}